\documentclass[fleqn,12pt,a4paper]{article}

\usepackage{amsthm}
\usepackage{amsmath}
\usepackage{amssymb}
\usepackage{amsopn}
\usepackage{latexsym}
\usepackage{setspace}
\usepackage{dsfont}
\usepackage{datetime}
\usepackage[a4paper,
top=0.95in, left=0.97in, right=0.97in, bottom=0.95in
]{geometry}
\usepackage[allnumber,warning,easyold,math]{easyeqn}
\usepackage[12pt]{moresize}
\usepackage{natbib}
\usepackage{hyperref}
\usepackage{afterpage}
\hypersetup{colorlinks=false,linkcolor=blue,
filecolor=magenta,urlcolor=blue,citecolor=blue,hidelinks}
\usepackage{fix-cm}
\usepackage[calcwidth]{titlesec}
\usepackage{graphicx}
\usepackage{sgame}
\usepackage[utf8]{inputenc}
\usepackage[english]{babel}
\usepackage{amssymb}
\usepackage{braket}
\usepackage{upgreek}
\usepackage[small]{caption}
\usepackage{subcaption}
\usepackage[ruled,vlined]{algorithm2e}
\usepackage{lipsum,fancyhdr,xcolor}
\usepackage{diagbox}
\usepackage{dsserif}
\usepackage{refcount}
\usepackage[dvipsnames]{xcolor}
\usepackage{soul}
\usepackage{tikz}
\usetikzlibrary{calc}
\usetikzlibrary{positioning}

\theoremstyle{plain}
\newtheorem{lem}{Lemma}
\newtheorem{thm}{Theorem}
\newtheorem{prp}{Proposition}

\newtheorem{cor}{Corollary}

\newtheorem{dfn}{Definition}

\newtheorem{obs}{Observation}

\newtheorem{example}{Example}



\newcommand{\R}{\mathbb{R}}

\newcommand{\winten}{\dot{\succsim}}
\newcommand{\sinten}{\dot{\succ}}
\newcommand{\iinten}{\dot{\sim}}
\newcommand{\wintenone}{\dot{\succsim}_1}

\newcommand{\wintentwo}{\dot{\succsim}_2}

\newcommand{\wintenthree}{\dot{\succsim}_3}

\newcommand{\wintenfour}{\dot{\succsim}_4}

\newcommand{\winteni}{\dot{\succsim}_i}

\newcommand{\wintenl}{\dot{\succsim}_l}
\newcommand{\sintenl}{\dot{\succ}_l}
\newcommand{\iintenl}{\dot{\sim}_l}

\begin{document}

\setcitestyle{authoryear}

\author{Georgios Gerasimou\thanks{\href{
mailto:
Georgios.Gerasimou@glasgow.ac.uk}{Georgios.Gerasimou@glasgow.ac.uk}.
This version: 20 July 2026. 
Earlier versions with different titles 
were presented in October 2023 at the 
Universities of Melbourne, Queensland and Glasgow; 
the 2022, 2024 Society for Social Choice 
and Welfare Conferences (online \& Paris);  
2024 Conference on Mechanism and 
Institution Design (Budapest); 
2024 EEA-ESEM (Rotterdam); 
2024 Hurwicz Workshop on Mechanism Design (Warsaw); 
2025 Coalition Theory Network Conference (Paris); 
2025 Durham Economic Theory Conference;   
2025 Economic Design Conference (Essex);
2025 World Congress of the Econometric Society (Seoul);
and 2026 BSE Forum (Barcelona).
The author thanks, alphabetically, 
Atila Abdulkadiro\u glu, Jean Baccelli, Yeon-Koo Che, 
Federico Echenique, Tore Ellingsen, Aytek Erdil, 
Marcelo A. Fernandez, Duarte Gon\c{c}alves, 
Peter J. Hammond, Herv\'{e} Moulin, Alex Nichifor, 
Jake Nebel, Viviane Pons, John Quah, Marek Pycia, Michael Richter 
and John Weymark for helpful conversations, 
literature pointers or expositional suggestions. 
First version: \href{https://arxiv.org/pdf/2011.04306v1}{November 2020}. 
}\\  \small University of Glasgow}

\renewcommand*{\thepage}{\footnotesize{\arabic{page}}}

\title{\vspace{-60pt}\LARGE\textbf{Efficient Allocation with 
\linebreak Ordinal Preference Intensities}}

\parindent=0pt

\date{}

\maketitle

\begin{abstract}
\noindent
Standard allocation methods that rely on ordinal preferences 
ignore how strongly agents value different improvements.
On the other hand, methods based on cardinal utilities 
require assumptions that are often considered too demanding. 
This paper studies the classic one-to-one assignment problem  
in the middle-ground environment of \textit{ordinal} preference 
intensities: agents rank alternatives 
as well as preference improvements, without necessarily   
being able to quantify these comparisons.
The paper introduces an interpersonal comparability 
assumption for this novel 
social-choice analytical setting. Building on it, 
it proposes and analyses the 
\textit{intensity-efficiency} criterion. This 
refines Pareto's criterion by 
discarding any allocation that is worse than another 
according to an intuitive intensity-dominance relation. 
The feasibility and applicability of this framework  
in matching problems is illustrated with two 
quadratic-time algorithms which, respectively, 
elicit ordinal intensity rankings and extend 
the classic Random Priority mechanism in 
a welfare-improving direction. 
\end{abstract}

\vspace{3pt}

\setcounter{page}{0}

\thispagestyle{empty}

\parindent=15pt

\begin{quotation}\footnotesize

\noindent \textit{``Suppose I am left 
with a ticket to a Mozart concert I am unable 
to attend and decide to 
give it to one of my closest friends. 
Which friend should I actually give it to? 
One thing I will surely consider 
in deciding this is which friend 
of mine would enjoy the concert \textnormal{most}.''}

\hfill John C. Harsanyi (\citeyear{harsanyi87})

\end{quotation}

\vspace{-4pt}

\begin{quotation}\footnotesize

\noindent \textit{``The problem I have with utilitarianism 
is not that it is excessively rational, 
but that the epistemological foundations are weak. 
My problem is: What are those objects
we are adding up? 
I have no objection to adding them 
up if there's something to add.''\footnotemark}

\hfill Kenneth J. Arrow 
(\citeyear{arrow-kelly87})
\end{quotation}

\footnotetext{Quoted in \cite{ellingsen}.}

\vfill

\pagebreak

\section{Introduction}

Distributional criteria for the assignment of goods 
to individuals can be evaluated jointly on the realism of their 
assumptions and the appeal---according to some justice principle---of 
their prescribed allocations. 
Arguably the most important such criterion is Pareto efficiency:
assuming only information on agents' preferences over
the relevant goods, its notion of optimality is defined by   
the requirement that no changes can make some individuals better off 
without hurting others. 
Yet it is well-known that among 
the typically many Pareto-efficient allocations 
there are often some which violate intuitive notions 
of distributive justice. 
Appreciation of these facts raises the question of whether 
Pareto's criterion might be improved on by 
considering alternative principles which, in addition to agents' 
preferences, also incorporate information on their 
\textit{preference intensities}. 
Economists have traditionally interpreted these 
to be synonymous to the existence of a \textit{cardinal} utility function 
that represents an agent's preferences over alternatives. 
Typically, this has been of the \citet{neumann_morgenstern} (vNM) kind, 
derived from expected-utility preferences over \textit{lotteries} over 
alternatives. Assuming that utility differences from such functions 
are interpersonally comparable, the classical utilitarian 
summation criterion, formalized in \cite{harsanyi55}, has been the 
benchmark distributional principle of the cardinal-welfarist tradition,
refining the set of Pareto-efficient allocations in 
intensity-inclusive directions. 

Despite the undeniable tractability and widespread application 
of this and related criteria in various domains of economic analysis, 
it is well-understood that %the specific 
cardinal utility functions emerging from expected-utility preferences over 
lotteries confound an individual's intensity comparisons with 
that person's attitudes toward risk
\citep{arrow51,arrow63,luce&raiffa,ellingsen,baccelli&mongin,sen17,alaoui-penta25,samuelson25}.
This fact, in turn, has generated concerns---such as the one 
expressed by Kenneth Arrow in the 
cited quotation---about the epistemic origins of such functions.
In addition, and no less importantly, when cardinal utility 
differences are used to define an agent's 
intensity comparisons, the person in question 
is portrayed as having a supernaturally high degree of precision in their 
capacity to quantify those intensities \citep{samuelson38restud,basu82}. 
Acknowledging the logical implication of this fact---namely,
that elicitation of agents' utilities in reality may be 
subject to ``noise''---diminishes a planner's confidence that application of 
utilitarian or related criteria in practical allocation 
problems is guaranteed to achieve the intended distributive outcome. In the 
distinct but related domain of evaluative voting problems, for example, 
where agents rate the candidates on numerical scales, 
\citet{baujard-etal18,baujard-igersheim-lebon21} provide
experimental evidence suggesting that the use of different scales 
leads to non-equivalent voting results.

%\subsection{Contribution}

This skepticism notwithstanding, intensity comparisons 
such as those in \citeauthor{harsanyi87}'s (\citeyear{harsanyi87}) 
opening quotation are common in everyday life and, in fact, do not 
presuppose the existence of interpersonally comparable utilities. 
They do, however, implicitly assume that:  
(i) agents can rank alternatives according to their preferences 
\textit{and} rank preference improvements according to the \textit{intensity} 
of these preferences; 
(ii) systematic interpersonal comparisons of such second-order data  
are possible. 
Given all the above, this paper's central question can
now be stated: \textit{``When agents cannot provide credible cardinal utilities, 
but can report which preference improvements feel larger than others, 
can a planner use that information to improve upon Pareto-efficient allocations?''}
We answer this question by focusing on the classic 
one-to-one assignment problem where 
finitely many indivisible goods must 
be assigned to as many agents. 

To address requirement (i) we assume that agents' preference intensity
orderings belong to the broad class of those that admit a generally
\textit{ordinal} numerical representation by means of a 
\textit{preference intensity function} \citep{gerasimou21}. 
In effect, this portrays each individual as being able to 
rank the above-mentioned preference improvements in a coherent way,
yet without necessarily being able to quantify them. 
This modeling approach is one possible way of dealing with the concern expressed 
earlier regarding cardinal utilities, and is in line with some 
contemporary views in cognitive psychology that are elegantly 
summarized in \citet*[p. 551]{vlaev-etal11} with the remark that 
\textit{``the perceptual system might be like a pan 
in balance, which responds by tipping to the left or right, depending
on which of two items is heavier, but provides no read-out of 
the absolute weight of either item.''}
To address requirement (ii), moreover, we employ an intuitive
normalization of agents' preference intensity functions and 
introduce an interpersonal comparability assumption that is suitable 
for this kind of ordinal intensity data. 
With this interpretational assumption in the background, finally, 
we propose and analyse a 
refinement of Pareto 
efficiency---\textit{intensity-efficiency}---that 
incorporates such data and discards allocations that are inferior 
to others in the light of this additional information, 
and according to an intuitive dominance relation. 

We illustrate both the interpersonal comparability 
assumption and the intensity-efficiency criterion 
with a very special example, which nevertheless helps  
make it clear from the outset how these ideas are new and 
distinct from those that underpin the utilitarian or 
\textit{Borda-scoring} \citep{borda,mwg,maskin25a} criteria. 
Imagine three football fans, 
labelled \textit{F1}, \textit{F2} and \textit{F3}, 
who are interested in watching, in person,  
one of three FIFA 2026 World Cup matches, labelled 
$a_1$, $a_2$ and $a_3$. 
Suppose that their valuations for the respective tickets 
are as shown in Table \ref{tab:valuations}.
Which ticket should go to which fan?
Both the Pareto and Borda criteria 
are tied between allocations 
$(a_1,a_2,a_3)$---which assigns $a_1$, $a_2$, $a_3$
to \textit{F1}, \textit{F2}, \textit{F3}, respectively---and 
$(a_2,a_1,a_3)$. 
Assuming that intra-personal differences in valuations 
are inter-personally comparable, 
the utilitarian criterion singles out allocation $(a_1,a_2,a_3)$.
By contrast, under the same assumption, the intensity-efficiency
criterion selects allocation $(a_2,a_1,a_3)$. 
The discrepancy is caused by the distinct ways in which 
the postulated utility differences are used by the two 
criteria. For the utilitarian one, the fact that \textit{F1}'s
difference of \$500 between $a_1$ and $a_2$ exceeds 
the respective difference of \$60 for \textit{F2} is the sole 
reason to break the tie in favour of the former. 
On the other hand, 
the intensity-efficiency 
criterion takes into account \textit{all} utility differences 
at the individual level in this special case where such 
differences are well-defined;
then considers how these are ranked 
(also at the individual level); 
and finally breaks the tie in favour of the fan for whom the 
utility difference between $a_1$ and $a_2$ is ranked higher 
in that person's intensity order. 
Thus, even though \textit{F1} (the possibly more affluent fan) 
has higher ticket valuations and valuation differences 
than \textit{F2}, ``levelling the ground'' 
between them by comparing the \textit{relative rankings} 
of their utility differences 
favours \textit{F2} in the assignment of the mutually preferred 
$a_1$ because, for that agent, the improvement from $a_2$ to $a_1$
is \textit{relatively} higher than the improvement from $a_3$ to $a_2$ 
in comparison to \textit{F1}.\footnote{This example is introduced 
to illustrate how the key 
new concepts that we introduce differ from existing ones. 
We do not take a stand in this paper on which line of reasoning may be 
``better'', and when, in this special case where intensities 
are defined by cardinal utilities. 
Although this is an interesting question, 
it is also one that lies outside the scope of this study
given the focus on what could be said and done 
when intensities are merely ordinal.}

\begin{table}[!htbp]
\centering
\arraycolsep=0pt\def\arraystretch{1.2}
\caption{Example valuations of indivisible items
$a_1$, $a_2$, $a_3$ by three agents.}
\begin{tabular}{|l|c|c|c|}
\hline
	  &\textit{F1}&\textit{F2}&\textit{F3} \\
\hline
$a_1$ & \$2,000 & \$100 & \$20 \\
\hline
$a_2$ & \$1,500 & \$40  & \$30 \\
\hline
$a_3$ & \$800   & \$20  & \$50 \\
\hline
\end{tabular}
\label{tab:valuations}
\end{table}

More generally, suppose a planner knows 
the agents' preference orders as well as their rankings over
preference improvements. The latter may have come from  
the agents' differences in willingness to pay, 
as in the above example. 
But, in general, we imagine them as being obtained from responses 
to more targeted questions such as ``Do you prefer 
$a$ to $b$ more than you prefer $c$ to $d$''?\footnote{\citet[Section 6 
\& Table I]{ellingsen} also promotes 
the use of data from responses to such questions 
toward eliciting information about agents' 
preference intensities and conducting social welfare analysis.} 
In the presence of such data, 
utilitarianism is inapplicable, while  
the Borda criterion is unresponsive by construction.
The interpersonal comparisons of ordinal intensities that 
we propose maintain the central idea that was displayed
in the example above. But, instead of comparing the 
relative ranking of utility differences 
(which may not be defined now), compares how an improvement from 
one alternative to another---or, synonymously,  
the \textit{intensity difference} between these
alternatives---is ranked by different agents, and interprets the 
better alternative as being preferred more intensely 
by the agent whose improvement lies higher 
in his/her intensity ordering relative to the other agents.
Intensity-efficiency builds on this interpersonal comparability 
criterion and requires that, other things equal, 
whenever two Pareto-efficient allocations assign the same two items to 
the same two agents but in a ``flipped'' way, and one 
allocation assigns the commonly
preferred item in every such pair to the agent who prefers it more, 
then that allocation should be favoured and the other one
discarded as \textit{intensity-dominated}.
Any Pareto-efficient allocation that is not dominated in this sense
is intensity-efficient.

Assuming strict preferences and strict rankings over improvements 
that need not be derived from ordered utility differences, 
in Theorem 1 we show that an intensity-efficient allocation 
exists under every profile when there are three agents and alternatives. 
Interestingly, however, 
existence is not guaranteed when there are more. 
The reason is that the underlying novel 
intensity-dominance relation over Pareto-efficient 
allocations (permutations) may be cyclic.
We then identify a property of intensity profiles that plays 
a central role in the existence of intensity-efficient allocations 
in the general case of $n$ agents and items.
This property imposes a \textit{homeo-monotonicity} structure within 
any \textit{monotonic} set of $m$ agents who have identical preferences 
over their $m$ top-ranked alternatives, namely a set where 
agents can also be ordered (possibly weakly)
in terms of how intensely they rank their preference improvements 
between any two of these items that appear consecutively in their 
preference list. Homeo-monotonicity---etymologically, the property of
increasing/decreasing in a similar fashion---simply requires this 
ordering over agents to not be violated when one also looks at how 
intensely agents rank items that appear \textit{non-consecutively} 
in their preference list.

For example, consider agents \textit{G1}, 
\textit{G2}, \textit{G3} and items $b_1$, $b_2$, $b_3$
over which the agents have the identical preferences 
$b_1\succ b_2 \succ b_3$; every other item is inferior to $b_3$; 
and \textit{G1} ranks the improvements from 
$b_3$ to $b_2$ and from $b_2$ to $b_1$ weakly 
higher than \textit{G2}, 
who in turn ranks them weakly higher than \textit{G3}. 
Homeo-monotonicity requires 
the improvement from $b_3$ to $b_1$ to be ranked weakly higher 
by \textit{G1} too, followed by \textit{G2} and then 
by \textit{G3}.
In Theorem \ref{thm:homeomonot-neces} 
we show that if preferences and intensities at a profile 
are ``polychotomous'' in the sense that 
all agents and items can be partitioned into 
monotonic sets and, possibly, an additional envy-free set where 
the remaining agents have distinct most preferred items, 
then, if an intensity-efficient allocation exists, there can be no  
monotonic set that is not homeo-monotonic 
whenever the underlying ordering over agents in 
that set is strict.
Conversely, in Theorem \ref{thm-homeomonot} we show 
that if a profile can be partitioned into 
homeo-monotonic sets and, 
possibly, an envy-free set, then it has an intensity-efficient 
allocation. Although the identification of a 
condition that fully characterizes the existence of intensity-efficient 
allocations is left as an open problem, based on our analysis 
and provided (counter-)examples we conjecture that this 
features some variation of homeo-monotonicity. 

In the last part of our analysis we turn to the practical 
application of these new welfare concepts. 
First, we show that, once agents' preferences are available, 
their ordinal intensities can be elicited in quadratic time
(Proposition \ref{prp:elicit}). 
We then illustrate how matching mechanisms can be constructed that 
assign to an ordinal intensity profile an allocation  
that could improve upon the Pareto-efficiency criterion in the 
intensity-dominance direction. 
Specifically, building on the matching-theoretic literature 
with one-sided preferences, we introduce an extension of the  
fundamental Random Priority (RP) mechanism that  
produces---in quadratic time in the number of items/agents---an
allocation which either coincides with or 
intensity-dominates the Pareto-efficient one 
that is found by RP (Proposition \ref{prp:complexity}).
Notably, such a welfare improvement is possible even in 
the very rare---as the large and unrestricted simulations 
that we report on in 
Section \ref{subsec:sims} suggest---cases where 
the input intensity profile does not have an 
intensity-efficient allocation. 
Although RP makes the elicitation of agents' preferences
incentive-compatible, we show that this does not carry over to 
the elicitation of agents' intensities under the extended algorithm. 
We connect this analysis to recent work 
in the related literature that started with 
\citet*{abdulkadirogluetal11}
(see also \citeauthor{miralles09}, \citeyear{miralles09};
\citeauthor{pycia11}, \citeyear{pycia11};
\citeauthor{featherstone-niederle16}, \citeyear{featherstone-niederle16})
and focuses on the trade-off between utilitarian welfare 
and incentive-compatibility that is relevant in a designer's choice 
between ordinal and cardinal matching mechanisms.

\subsection*{Related Literature}

This paper's contribution lies at the 
intersection of welfare economics, social choice and matching theory. 
Similar to existing cardinal-welfarist refinements 
of Pareto-efficiency such as those based on classical or 
relative utilitarianism 
\citep{harsanyi55,dhillon-mertens99,harvey99,nebel24,weymark05}, 
Nash social welfare \citep{nash50,luce&raiffa,moulin04} and others,
the proposed refinement is inclusive of information that 
goes beyond agents' preferences 
toward also reflecting their intensities in the allocation process. 
Unlike the former refinements, the hereby proposed one is the first 
that applies in the general class of ordinal intensity profiles that 
comprise preferences and rankings over preference improvements. 
Furthermore, it is formulated in a novel analytical environment 
that builds on a new interpersonal-comparability assumption. 
As the discussion around the example of Table \ref{tab:valuations} 
shows, even in the very special cases where 
agents' intensities are defined in terms of cardinal utility 
differences, the way in which these are interpersonally compared 
differs from how they are so under the above cardinal-welfarist 
criteria.

Intensity-efficient allocations and their domain of application 
are also distinct from those dictated by the Borda scoring rule 
\citep{borda,young74,maskin25a}. 
Although this rule is defined by a particular normalization 
of agents' \textit{ordinal} utility functions over alternatives, 
in practice it is often used with a cardinal interpretation 
due to the fact that ordering the utility differences induced by this 
normalization defines the special class of intensity relations 
where every agent prefers two alternatives that are ranked $k$ 
places apart \textit{exactly as much} 
as she prefers \textit{any} two other alternatives that are ranked 
$k$ places apart. The class of ordinal intensity relations 
that we consider does not impose this restriction. Yet at the same time,   
for the main results of this paper we assume a strictness condition 
on agents' intensities which rules out such ``linear'' intensity 
profiles.
Intensity-efficient allocations are also fundamentally 
different from \textit{ordinally efficient} (random) allocations
in the sense of \cite{bogomolnaia-moulin01}.
Indeed, a random allocation is said to have this 
property at some strict preference profile if there is 
no other random allocation that stochastically dominates it under
the cumulative density functions that are defined relative to the 
agents' preference orderings.\footnote{That is, 
a random allocation is ordinally efficient if,  
for every agent and every position $r$ in that 
agent's preference order, no other random allocation always gives a 
weakly higher probability to that agent receiving one of their 
$r$ most preferred items than the former allocation does, 
with this inequality being strict for some agent and position $r$.}

In addition to \cite*{abdulkadirogluetal11}, finally, 
several studies in matching theory---both with one- and two-sided
preferences---have taken agents' interpersonally comparable 
vNM cardinal utilities 
\citep*{hylland&zeckhauser,budish11,lee&yariv18,ortoleva-safonov-yariv} 
or quasi-linear preferences in willingness to pay 
\citep*{che&gale&kim13} as inputs 
in the allocation process toward maximizing utilitarian welfare. 
Our study shares the same goal as those above, namely to improve upon 
the distributive-justice properties of Pareto-efficiency in the final matching. 
Its distinguishing point is that it pursues this in the unexplored 
environment where only ordinal intensity comparisons are generally 
available to the designer.
While this environment is less tractable for the 
modelling analyst at present, it is cognitively less demanding 
for the economic agent and, as such, we believe worthy 
of further investigation.

\section{Preliminaries}

By $A:=\{a_1,\ldots,a_n\}$ and $N:=\{1,\ldots,n\}$ 
we denote, respectively, the finite sets of choice alternatives 
and agents. To simplify the assignment problem under study we 
assume that these sets have equal sizes. 
When subscripts are unnecessary we also write $a,b\in A$.
The \textit{preference intensity relation} of agent $l\leq n$ 
on $A$ is denoted by $\wintenl$, with $\sintenl$ and $\iintenl$ 
its asymmetric and symmetric parts. 
In line with extensive-measurement theory \citep{krantzetal,pfanzagl71,roberts79}, 
this is a quaternary relation on $A$ or, equivalently, 
a binary relation on $A\times A$. 
The statement $(a,b)\, \wintenl\, (c,d)$ can be read as 
``$a$ is preferred to $b$ at least as much as $c$ is to $d$''
when the first option in each pair is the (possibly weakly) 
preferred one at that pair, and as ``$b$ is preferred to 
$a$ no more than $d$ 
is preferred to $c$'' when the converse 
is true (naturally, $(a,b)\wintenl(c,d)$ 
will also hold when $a$ is preferred to 
$b$ and $d$ is preferred to $c$). 
When no ambiguity arises, 
we will interpret $(a,b)\, \wintenl\, (c,d)$ 
more succinctly as suggesting that 
the \textit{intensity difference} between
$a$ and $b$ is weakly larger than that between $c$ and $d$,
or that the \textit{improvement} from $b$ to $a$
is bigger than the one from $d$ to $c$.
In line with these interpretations, and the literature, 
agent $l$'s binary preference relation $\succsim_l$ on $A$ 
is derived from their intensity relation $\wintenl$ 
by $a\succsim_l b \Leftrightarrow (a,b)\, \wintenl\, (b,a)$.
We will refer to $\succsim_l$ as the preference relation that is 
\textit{induced} by $\wintenl$. 
As usual, the asymmetric and symmetric parts of $\succsim_l$ 
are denoted by $\succ_l$ and $\sim_l$.
Under the structure that $\wintenl$ will be endowed with below,
both this relation and its induced $\succsim_l$ are weak orders
on their respective domains. 

For $a,b\in A$, the \textit{intensity-equivalence class} 
of $(a,b)\in A\times A$ under $\wintenl$ is defined 
by 
\begin{eqnarray}
[a,b] & := &
\{(a',b')\in A\times A: (a',b')\, \iintenl\, (a,b)\}
\end{eqnarray}
Furthermore, the quotient set of $A\times A$ under $\iintenl$
is defined by 
\begin{eqnarray}
(A\times A)_{\iintenl} & := & \{[a,b]: [a,b] 
\text{ is an intensity-equivalence class under } \iintenl\}
\end{eqnarray}

The definitions that follow omit universal quantifiers, 
yet they should be understood as imposing conditions that 
apply to all objects in the respective domains.
%\begin{dfn}	\label{UD-dfn}
A relation $\wintenl$ has a
\textit{utility-difference representation} 
if there is a function $u_l:A\rightarrow \mathbb{R}$
that satisfies
\begin{eqnarray}
\label{UD-eq}	
(a,b)\, \wintenl\, (c,d) & 
\Longleftrightarrow & 
u_l(a) - u_l(b) \geq u_l(c) - u_l(d)
\end{eqnarray}
%\end{dfn}
In an influential article, psychologist S. S. \citeauthor{stevens46}
(\citeyear{stevens46}) distinguished between four scales 
of measurement that are afforded by some numerical assignment  
on a set of items: \textit{nominal} (qualitative/categorical measurement);
\textit{ordinal} (invariance up to a strictly increasing transformation); 
\textit{interval} (invariance up to positive affine transformation); 
and \textit{ratio} (invariance up to a positive linear transformation).
Utility-difference representations were added to this list 
by psychologist C. H. \citeauthor{coombs50} (\citeyear{coombs50}) as those that 
correspond to \textit{ordered metric scale} measurement
(see also \citeauthor{ellingsen}, \citeyear{ellingsen}).
These representations---analysed in \cite{scott&suppes} 
and \cite{scott64}---are more general than neoclassical 
cardinal utility representations
because $u_l$ in the latter case is an interval scale
whereas in the former it \textit{``falls logically between 
an interval scale and an ordinal scale''} (p. 145).\footnote{
More specifically, it has the complex uniqueness property of additive 
representations on finite sets, articulated in  
\citet[Theorem 2, p. 431]{krantzetal}. \cite{baccelli24} has recently 
clarified, however, that 
every ordinal utility representation defines a certain subset of  
utility-difference comparisons whose order is, in fact, preserved
by all ordinal transformations of the original representation.}

In \cite{gerasimou21}, this author recently
proposed and characterized the class of 
\textit{ordinal} intensity relations 
that admit the following more general numerical representation.
More specifically, 
%\begin{dfn}
a relation $\wintenl$ 
is representable by a 
\textit{preference intensity function} if 
there is a mapping
$s_l:A\times A\rightarrow \mathbb{R}$,
unique up to a strictly increasing transformation, 
such that
\begin{eqnarray}
\label{PIF1} 	
(a,b)\,\wintenl\, (c,d) & 
\Longleftrightarrow & s_l(a,b)\geq s_l(c,d)\\	
\label{PIF2} 	
(a,b)\,\wintenl\, (b,a) & \Longleftrightarrow &
s_l(a,b)\geq s_l(e,e) \geq s_l(b,a)\\
\label{PIF3} 	
\min\{s_l(a,b), s_l(b,c)\} \geq s_l(e,e) & 
\Longrightarrow & s_l(a,c) \geq 
\max\{s_l(a,b), s_l(b,c)\},
\end{eqnarray}
with a strict inequality on the left part 
of \eqref{PIF3} implying a strict inequality on the right.
%\end{dfn}
In words, $s_l$ represents the agent's 
intensity weak ordering [\eqref{PIF1}] and
the preferences induced by it [\eqref{PIF2}], 
further ensuring that these preferences are themselves 
weakly ordered and intensities are monotonically 
increasing in the preference ordering [\eqref{PIF3}].
This ordinal model nests the (pseudo-)cardinal 
utility-difference model in the special case where 
\textit{``lateral consistency''} [\eqref{PIF3}] 
is strengthened to \textit{``additivity''} [\eqref{PIF4}]:

\begin{lem}
\label{lem:sincov}
A relation $\wintenl$ that is represented by a preference intensity function
$s_l$ is utility-difference representable if and only if a strictly 
increasing transformation $\widetilde{s}_l$ of $s_l$ satisfies 
\begin{eqnarray}
\label{PIF4} \widetilde{s}_l(a,c) & = & 
\widetilde{s}_l(a,b) + \widetilde{s}_l(b,c)
\end{eqnarray}
\end{lem}
\noindent 
Unless indicated otherwise, proofs appear in the Appendix.
Lemma \ref{lem:sincov} states that an intensity ordering 
$\wintenl$ that is ordinally representable in the sense of 
\eqref{PIF1}--\eqref{PIF3} is 
also representable in the utility-difference sense of 
\eqref{UD-eq} by some (pseudo-)cardinal utility function $u_l$
if and only if \textit{some} intensity function that represents
$\wintenl$ can be constructed so as to also satisfy the 
additivity condition \eqref{PIF4}. 
That such a representation is not cardinal in general 
can easily be seen with an example: suppose $u_l$ and $u'_l$
are defined on $A=\{a,b,c\}$ by $u_l(a)=10$, $u_l(b)=6$, $u_l(c)=0$
and $u'_l(a)=10$, $u'(b)=7$ and $u'_l(c)=0$. Both functions
represent the intensity order $(a,c)\sintenl (b,c)\sintenl(a,b)
\sintenl(a,a)\iintenl(b,b)\iintenl(c,c)\sintenl(b,a)\sintenl (c,b)
\sintenl(c,a)$; and both have the same minimum and maximum
values. Yet they are not related by a positive affine transformation.

Next, we will say that a preference intensity function $s_l$ 
represents $\wintenl$ \textit{canonically} if 
\begin{eqnarray}
s_l(A\times A) & = & \{-k,-k+1,\ldots,-1,0,1,\ldots,k-1,k\},
\label{canonical}
\end{eqnarray}
where $k$ is the number of intensity equivalence classes
$[a_i,a_j]\in (A\times A)_{\iinten_l}$ 
such that $a_i\succ_l a_j$.
%\end{dfn}
Henceforth, $s_l$ will denote the canonical intensity
function that represents $\wintenl$. 
As will be clarified later, the usefulness of this normalization 
for our purposes lies in the fact that the integer value it associates 
with a pair reveals the rank position of that 
pair in the agent's intensity ordering.

\begin{lem}
\label{lem:canonical}
If $\wintenl$ can be represented by a preference intensity function, 
then it admits a canonical such representation.
\end{lem}

We assume throughout that every $\wintenl$ is representable 
by some $s_l$; hence, by Lemma \ref{lem:canonical} canonically so.
Furthermore, our subsequent analysis will 
impose the following additional property on $\wintenl$:

\vspace{5pt}

\noindent 
\textbf{Strictness}.
\textit{$(a,b)\, \mathord{\not\mathrel{\iintenl}}\,(c,d)$ for all 
distinct off-diagonal pairs of items $(a,b)$ and $(c,d)$.}

\vspace{5pt}

\noindent This condition rules out the possibility of the agent 
preferring $a$ to $b$ \textit{exactly} as much as they prefer $c$ to $d$ 
for distinct pairs of distinct alternatives $(a,b)$ and $(c,d)$.
It is therefore analogous to the standard 
preference-strictness postulate which, in fact, it implies. 
As we elaborate in Section 3, 
strict intensities are useful for our purposes because they ensure that 
the canonical intensity functions of all agents are onto the same set 
$\{-k,\ldots,-2,-1,0,1,2,\ldots,k\}$,
where $k\equiv {n\choose 2}$. 
In particular, the only non-trivial intensity-equivalence class
when $\wintenl$ satisfies Strictness
is the diagonal subset of $A\times A$, with every pair $(a,a)$ 
in this set being mapped to 0 by the canonical 
representation of $\wintenl$.
Hence, Strictness also rules out the kinds of ``linear''
intensity relations that are implicit in the cardinal 
interpretations of Borda-normalized utility functions, 
where the agent is portrayed as considering 
the intensity differences between any two 
consecutively-ranked items to be equivalent.
Just like the assumption of strict preferences that is so often 
invoked in social-choice and matching-theoretic analyses, 
the assumption of strict intensities is demanding 
and possibly unrealistic in certain contexts. 
It does, however, enable us to introduce this paper's 
key new ideas and results in a relatively tractable environment 
which encompasses the more familiar one where the principal
assumption is that all agents' preferences are strict
\citep{abdulkadiroglu&sonmez13,echenique-immorlica-vazirani-23a}.

Because the combined effect of the Strictness assumption and 
the---by Lemma \ref{lem:canonical}, without loss---canonical 
normalization is important in what follows, 
we clarify the following:

\noindent 
\textit{If $s_l$ is agent $l$'s canonical intensity function
and there are items $a,b,c,d$ and a positive 
integer $k$ such that $s_l(a,b)=k\,s_l(c,d)$,  
\textbf{this does not imply} that the intensity difference between 
$a$ and $b$ is $k$ times the difference between $c$ and $d$.} 

\noindent 
This is evident upon a moment's reflection, since the values 
of a canonical $s_l$ at different pairs merely capture the 
\textit{ranks} of those pairs in the intensity ordering $\wintenl$. 
We emphasise it in order to dispel potential misunderstandings 
that may spill over to the assumption that is introduced in 
the next section.

\section{Interpersonal Comparability of Ordinal Intensities}

We denote by $\mathcal{I}$ the class
of intensity relations that are representable by 
a preference intensity function and are also strict in the sense 
described previously. 
Furthermore, we let $S\, =\, (\winten_1,\cdots,\winten_n)$  
stand for a (strict) \textit{intensity profile} where 
$\wintenl\in \mathcal{I}$ for all $l\in N$, and 
write $\mathcal{S}$ for the collection of all such profiles. 
By $P_S\, =\, (\succ_1,\ldots,\succ_n)$ we denote the (strict) 
\textit{preference profile} that is induced by $S\in\mathcal{S}$,
and by $\mathcal{P}$ the collection of all strict preference profiles.
For any $S\in\mathcal{S}$, $s^S=(s_1^S,\ldots,s_n^S)$ 
is understood as the profile of canonical preference intensity functions 
that represent the agents' preferences and intensities, as specified  
in $S$. 
When no confusion arises we simply write $s$ and $s_l$ instead 
of $s^S$ and $s_l^S$, respectively.
In the sequel, and unless otherwise noted, 
``profile'' refers to a strict intensity profile, 
i.e. some $S\in\mathcal{S}$. 

\vspace{5pt}

\noindent \textbf{Assumption}\\
\textit{Given a profile that is represented canonically as 
in $s=(s_1,\ldots,s_n)$, the statement}
\begin{eqnarray}
\label{comparison}
s_l(a,b) & > & s_m(c,d) \; > \; 0 
\end{eqnarray}
\textit{is interpreted as suggesting that agent $l$ prefers $a$ to $b$ 
\textnormal{more} than agent $m$ prefers $c$ to $d$.}

\vspace{5pt}

Towards motivating this novel assumption---whose relevance, 
we stress, is interpretational---we first recall 
that an individual's intensity comparisons 
are not assumed to be quantifiable beyond the level 
of an ordinal ranking. 
Yet \textit{some} information about the generally 
different welfare effects of enjoying $a$ or $b$ vs. 
$c$ or $d$ \textit{is} available. 
Suppose, in particular, that 
$a\succ_l b\succ_l c \succ_l d$, and agent $l$ 
prefers $a$ to $b$ more than she does $c$ to $d$. 
This is equivalent to saying that her intensity difference 
at pair $(a,b)$ is greater than at $(c,d)$.
Consider agent $m$ next, and suppose for simplicity that the 
same applies to him too: $s_m(a,b)>s_m(c,d)$.  
Given this, and given also that both agents have preferences 
and intensities over the same set of alternatives, how might 
\eqref{comparison} be interpreted? 
We know that the intensity difference within pair $(a,b)$ is ranked 
higher by agent $l$ in her intensity ordering
than the corresponding difference within pair $(c,d)$ is ranked 
by agent $m$ in his.
By the Strictness assumption, moreover, $l$ and $m$ have  
the \textit{same} number of possible rank positions in their respective 
intensity orderings, which pairs $(a,b)$ and $(c,d)$ could possibly occupy. 
In other words, the agents' preference-improvement comparisons 
are on level ground because their ordinal intensity scales 
coincide: 
if $(a,b)$ and $(c,d)$ are ranked 5th and 8th in $l$'s and $m$'s scales,
respectively, then they are ranked 5th and 8th out of the 
\textit{same} $k$ possible positions.

Although distinct in important respects, 
this logic resembles the way scoring rules
such as the normalization of ordinal 
utilities in the Borda count \citep[see \eqref{BO-0} below]{borda} 
or scoring-like rules such as  
the relative-utilitarian \citep{dhillon-mertens99} normalization of 
cardinal utilities\footnote{See \cite{d'Aspremont2018} for such 
a perspective on relative utilitarianism.} 
perform interpersonal comparisons in 
their respective environments. 
In particular, neither of these social welfare functions 
formally requires interpersonal utility comparisons in order 
for the respective social outcome to be immune to permissible 
transformations of the agents' utility functions.\footnote{See
\cite{baccelli23} for an especially clear recent discussion.}
Clearly, this is so because both welfare criteria are defined
in terms of specific normalizations which reduce, respectively,  
any ordinally or cardinally equivalent utility function that represents
an agent's preferences into the \textit{same} normalized ordinal 
or cardinal utility function. 
At the same time, however, 
\textit{applications} of the Borda and relative-utilitarian 
normalizations \textit{do} introduce interpersonal utility comparisons.
We elaborate on this in the case of the Borda method, 
which is conceptually closer to our setting. 
This takes agent $l$'s preference relation $\succ_l$
and defines the \textit{Borda score} of every item $a$ in $A$ by 
\begin{eqnarray}
\label{BO-0}
B_l(a) & := & |\{a'\in A: a\succ_l a'\}|,
\end{eqnarray} 
i.e. by the number of alternatives that $l$ considers inferior to $a$.
Once these normalised utility functions are constructed for all agents, 
the Borda social welfare function maps any such profile to a social ordering 
that prioritises items with higher aggregate Borda scores. 
Although interpersonal comparability assumptions are not part of the axioms
that characterize the Borda method, the motivation for its very existence
does implicitly invoke such comparisons. This is articulated clearly
in \citet[f.17]{maskin25a}, as follows: 
\textit{``Notice that the Borda count implies a particular way of 
making interpersonal comparisons, e.g., if individual 1 ranks $x$ 
two positions above $y$, that preference is exactly canceled
by two individuals who rank $y$ one position above $x$. 
Observe, however, that none of our axioms 
speaks directly to interpersonal comparisons at all; 
such comparisons are an emergent property of the joint 
imposition of the axioms''.}

One may now observe both the analogy and the qualitative differences 
between the canonically-normalized values that possibly different 
\textit{pairs of items} may occupy in the \textit{intensity orderings} 
of different agents and the possibly distinct positions 
(equivalently, Borda scores) 
of \textit{items} across agents's \textit{preference 
orderings}: 
unlike the additional 
cancellations that are imposed by the 
aggregation of Borda scores under that social welfare
function, the Assumption merely requires that an ordinal 
comparison be made about agent $l$'s intensity difference 
between items $a$ and $b$ against agent $m$'s such difference 
between $c$ and $d$.
As Section 2 hopefully makes clear, moreover, 
the values $s_l(a,b)$ and $s_m(c,d)$ themselves 
\textit{cannot} be interpreted as emerging---via an additive 
or multiplicative operation---from some underlying utility function 
(whether the Borda-normalized ones, $B_l$, $B_m$, or others), 
nor from the values of $s_l$ or $s_m$ at other pairs. 
Put differently, although the Assumption introduces interpersonal 
comparisons of scores, these scores: (i) reflect the rankings of 
\textit{pairs} of items (equivalently, of intensity differences
between items); (ii) are not derivable from some utility function
over items; and (iii) do not cancel themselves out interpersonally.

The question now emerges: 
taking as given that it is desirable in the first place for the 
information contained in the different agents' ordinal intensity 
orderings to be reflected in the allocation process, 
should these be treated equally by the planner?
While equal treatment may or may not be an appropriate approach 
to follow in practice, depending on what else is known about agents 
and the problem at hand, since the intensity orderings are assumed here 
to encode \textit{all} available welfare-relevant information, accepting their relevance
but treating them in any way other than equal would call for a justification that 
appears elusive. 
In light of all the above, we view the suggested interpretation of 
\eqref{comparison} as a reasonable basis for making interpersonal 
comparisons in this informational environment.

Another relevant question arises ---now on the analytical side: 
in the special case where the ordinal intensity ordering
of every agent is defined by how the pairwise differences of 
a cardinal utility function are ordered, and these differences 
in turn define a strict intensity profile, 
does interpersonal comparability of ordinal intensities 
in the sense of our Assumption reduce to the 
interpersonal comparability of utility differences that is 
required by \textit{utilitarian} aggregation? 
As the Table \ref{tab:valuations} example from the Introduction 
shows, the answer is \textit{no}. We show this below
using the formal notation that's been laid out since.

\begin{example}
\label{exm:discrepancy}
Let $n=3$ and consider the utility profile 
\begin{eqnarray*}
	U & := & \big(\underbrace{(2000,1500,800)}_{u_1},\ 
	\underbrace{(100,40,20)}_{u_2},\ 
	\underbrace{(20,30,50)}_{u_3}\bigr) 
\end{eqnarray*}
over $A=\{a_1,a_2,a_3\}$, defined from the valuations in 
Table \ref{tab:valuations}.
It is readily seen that the intensity profile 
defined by the intra-personal utility differences in $U$
is strict.
Also, defining $\sintenl$ by $(a_i,a_j)\sintenl (a_k,a_m)$ 
$\Leftrightarrow$ $u_l(a_i)-u_l(a_j) > u_l(a_k)-u_l(a_m)$, we observe
that $(a_2,a_3)\, \sinten_1\, (a_1,a_2)$ and 
$(a_1,a_2)\, \sinten_2\, (a_2,a_3)$---hence 
$s_2(a_1,a_2)>s_1(a_1,a_2)$---while 
$u_1(a_1)-u_1(a_2)>u_2(a_1)-u_2(a_2)$ is also true.
\hfill$\blacklozenge$
\end{example}

Thus, even if there is some ordinal transformation 
$\widetilde{s}_i$ of the 
canonical intensity function $s_i$ such that 
$\widetilde{s}_i(a,b)\equiv u_i(a)-u_i(b)$ for a cardinally 
unique $u_i$ [see~\eqref{PIF4}], the above clarifies that
\eqref{comparison} neither implies nor 
is implied by the inequality $u_l(a)-u_l(b)>u_m(c)-u_m(d)$.
For this reason, and in light of the preceding discussion, 
in such rich special environments we must generally distinguish 
between the \textit{relative intensity difference} that is captured 
by $s_l(a,b)>s_m(c,d)$ and the \textit{absolute intensity difference} 
that is captured by the above interpersonal 
utility-difference inequality,
which is preserved under any positive affine transformation of agents'
utilities that involves a common multiplicative factor (i.e. the 
\textit{Cardinal Unit Comparability} informational basis ---see
\cite{baccelli23} and references therein).
Echoing what was already mentioned in the Introduction, 
the general discrepancy in the normative 
conclusions that can be reached in these informationally 
rich---but, from this study's point of view, very special---cardinal 
environments where both relative and absolute intensity
comparisons can be made across individuals is of independent 
interest, economically as well as philosophically.
Cardinal environments not being the focus of this study, however, 
such an analysis will not be pursued here.

We conclude this section by stressing that the concepts 
introduced in the sequel rely on the position 
where a pair of alternatives lies within an agent's intensity ranking. 
Thus, by introducing canonically normalized intensity functions  
at this early stage we assign a specific name and piece of notation 
to these score-like positions. 
Crucially, this means that the welfare concepts of the next section are invariant 
with respect to arbitrary monotonic transformations of any intensity functions 
that one may choose to represent the agents' ordinal intensities by, 
because---similar to the Borda and relative utilitarian functions---these 
are automatically translated into canonically normalized values of ordinal 
intensity functions that are onto the same range across all agents.

\section{Intensity-Dominance and Intensity-Efficiency}

\subsection{Definitions and Existence}

Since the number of agents and items are assumed to coincide, 
an \textit{allocation} of the $n$ goods is a permutation on $A$. 
The set of all allocations is denoted by $\mathcal{A}$.
As was discussed in the Introduction, 
we are interested in the assignment of 
the $n$ objects in $A$ to the $n$ agents in $N$ in a 
way that satisfies Pareto-efficiency, but also improves upon it 
by delivering a narrower set of acceptable allocations that 
are appealing from a distributive-justice perspective 
in the present analytical environment of 
ordinal preference intensities.
To this end, we proceed with the introduction of 
the following novel notions of dominance and efficiency.

\begin{dfn}\label{intensity-dominance}
Given allocations $x$ and $y$, the former 
\textnormal{intensity-dominates} the latter 
at a profile 
if for every pair of agents $(i,j)$ such that 
$(x_i,x_j) = (y_j,y_i)=(a,b)$  it holds that
$s_{i}(a,b) \geq s_{j}(a,b)$,  
and the inequality is strict for some pair. 
\end{dfn}

\begin{dfn}
An allocation is \textnormal{intensity-efficient} at a profile 
if it is Pareto-efficient at the induced preference profile 
and is not intensity-dominated by another Pareto-efficient allocation. 
It is \textnormal{intensity-dominant} if it intensity-dominates 
every other Pareto-efficient allocation.
\end{dfn}

Suppose $x$ and $y$ are Pareto-efficient allocations that differ only 
in that, for some pairs of agents, both assign the same two items
within every such pair but in opposite ways. For example, 
$x$ allocates $a$, $b$ to agents 1, 2 and $c$, $d$ to agents 3, 4,
respectively, whereas $y$ allocates $b$, $a$ and $d$, $c$ to them.
Then, $x$ intensity-dominates $y$ if, in every such pair, it allocates 
the commonly preferred item to the agent who prefers it at least as
much as the other agent, and in at least one pair the agent who 
receives that item prefers it strictly more than the other agent.
Like Pareto-dominance, intensity-dominance is an incomplete binary 
relation over allocations. There are two (distinct but combinable)
ways in which this criterion may be unable to rank two allocations. 
The first is when no pairs of the kind postulated in the definition 
exist. For example, $(a,b,c)$ and $(c,a,b)$ are incomparable because 
of that. The second is when intensity-dominance across different pairs 
point in opposite directions and favour distinct allocations.
For example, $(a,b,c,d)$ and $(b,a,d,c)$ when $s_1(a,b)>s_2(a,b)$
and $s_4(c,d)>s_3(c,d)$ cannot be ranked because the first 
inequality favours the former allocation while the second favours
the latter. A Pareto-efficient allocation is therefore 
intensity-efficient at some profile if it either dominates or 
is not dominated by another Pareto-efficient allocation.
If it dominates every such allocation, it is intensity-dominant.  

With the following example we illustrate all three concepts 
in their full generality, and also highlight the potentially 
significant refinements on Pareto efficiency---with corresponding
welfare gains---that can be effected by intensity-efficiency.

\begin{example}
Let $A:=\{a_1,a_2,a_3,a_4\}$ and 
$S=(\wintenone,\wintentwo,\wintenthree,\wintenfour)$ 
be such that 

$$
\begin{array}{ccccccc}
s_1(a_1,a_4)=6&&s_2(a_4,a_1)=6&&s_3(a_1,a_4)=6&&s_4(a_4,a_1)=6 \\
s_1(a_2,a_4)=5&&s_2(a_4,a_2)=5&&s_3(a_1,a_3)=5&&s_4(a_3,a_1)=5 \\
s_1(a_1,a_3)=4&&\color{blue}s_2(a_4,a_3)=4&&\color{red}s_3(a_1,a_2)=4 &&s_4(a_4,a_2)=4 \\
s_1(a_2,a_3)=3&&s_2(a_3,a_1)=3&&s_3(a_2,a_4)=3&&s_4(a_3,a_2)=3 \\
\color{red}s_1(a_1,a_2)=2&&s_2(a_3,a_2)=2&&s_3(a_3,a_4)=2&&s_4(a_2,a_1)=2 \\
s_1(a_3,a_4)=1&&s_2(a_2,a_4)=1&&s_3(a_2,a_3)=1&&\color{blue}s_4(a_4,a_3)=1\\
\end{array}
$$ 
Agents 1 and 3, and agents 2 and 4, have identical preferences
but different intensities\footnote{We note without proof 
that $\winten_1$ does not admit a utility-difference representation
in the sense of \eqref{UD-eq}.}:
$$
\begin{array}{ccccccc}
a_1 & \succ_1 & a_2 & \succ_1 & a_3 & \succ_1 & a_4\\
a_4 & \succ_2 & a_3 & \succ_2 & a_2 & \succ_2 & a_1\\
a_1 & \succ_3 & a_2 & \succ_3 & a_3 & \succ_3 & a_4\\
a_4 & \succ_4 & a_3 & \succ_4 & a_2 & \succ_4 & a_1
\end{array}
$$
Hence, the Pareto-efficient allocations corresponding to 
induced preference profile $P_S$ are:

\centerline{
$\underbrace{(\color{red}a_1,\color{blue}a_3,
\color{red}a_2,\color{blue}a_4\color{black})}_w$,\; 
$\underbrace{(\color{red}a_1,\color{blue}a_4,\color{red}a_2,
\color{blue}a_3\color{black})}_x$,\; $\underbrace{(\color{red}a_2,
\color{blue}a_3,\color{red}a_1,\color{blue}a_4\color{black})}_y$,\; 
$\underbrace{(\color{red}a_2,\color{blue}a_4,\color{red}a_1,
\color{blue}a_3\color{black})}_z$}
\noindent
Given these preferences, and the fact that 
the two pairs of agents order the four items in opposite ways, 
these allocations alternate their assignments of 
$a_1$, $a_2$ to agents 1, 3, and of $a_3$, $a_4$ to agents 2, 4.
Invoking the intensity-dominance relation over Pareto-efficient
allocations using the information in the canonical profile 
$s$ presented above, we observe that the second agent 
prefers $a_4$ to $a_3$ more than the fourth agent, 
and that the third prefers $a_1$ to $a_2$ more than the first.
This makes $z$ the unique intensity-efficient allocation here, 
which, in fact, is also intensity-dominant.
Furthermore, allocations $x$ and $y$ are incomparable 
by dominance because the two possible intensity-based swaps 
favour distinct allocations. Finally, both $x$ and $y$
intensity-dominate $w$. \hfill $\blacklozenge$
\end{example}

\begin{thm}\label{thm:n=3}
Every profile has an intensity-efficient allocation 
when there are three agents and items. 
This is not true in general with four or more.
\end{thm}

\begin{table}[!htbp]
\centering
\caption{Example profile that has no intensity-efficient allocation.}
\arraycolsep=0pt\def\arraystretch{1.2}
\small
\begin{tabular}{c|cccc}
\backslashbox{\small\hspace{-10pt}$s_i(a,b)=$}{\hspace{-10pt}\small$i=$}
&1	
&2
&3
&4 \\
\hline
6 &$(a,d)$ &$(a,d)$ &$(a,d)$ &$(a,d)$ \\
5 &$(b,d)$ &\color{red}$(a,c)$ &\color{red}$(a,c)$ &$(b,d)$ \\
4 &\color{red}$(a,c)$ &$(b,d)$ &$(b,d)$ &\color{red}$(a,c)$ \\
3 &\color{blue}$(b,c)$ &\color{blue}$(b,c)$ &\color{blue}$(b,c)$ &\color{blue}$(b,c)$ \\
2 &\color{magenta}$(a,b)$ &$(c,d)$ &$(c,d)$ &\color{magenta}$(a,b)$ \\
1 &$(c,d)$ &\color{magenta}$(a,b)$ &\color{magenta}$(a,b)$ &$(c,d)$ \\
\end{tabular}
\label{tab:counterexample4}
\end{table}

We explain why non-existence is a potential issue when $n\geq 4$, 
as this will be useful for our subsequent analysis.
To this end, consider the example profile 
on $A=\{a,b,c,d\}$ whose canonical representation is shown 
in Table \ref{tab:counterexample4}.
Observe that $s_1=s_4$, $s_2=s_3$ and 
$a \succ_i b \succ_i c \succ_i d$, $i=1,\ldots,4$. 
This implies that all 24 possible allocations, defined and listed below, 
are Pareto efficient:
$$
\begin{array}{llllllllllllllllll}
x^1&=&(a,b,c,d)&&&x^2&=&(a,b,d,c)&&&x^3&=&(a,c,b,d) &&& x^4&=&(a,c,d,b)\\
x^5&=&(a,d,b,c)&&&x^6&=&(a,d,c,b)&&&x^7&=&(b,a,c,d) &&& x^8& = &(b,a,d,c)\\
x^9&=&(b,c,a,d)&&&x^{10}&=&(b,c,d,a)&&& x^{11}&=&(b,d,a,c)&&& x^{12}&=&(b,d,c,a)\\
x^{13}&=&(c,a,b,d)&&&x^{14}&=&(c,a,d,b)&&&x^{15}&=&(c,b,a,d)&&&x^{16}&=&(c,b,d,a)\\
x^{17}&=&(c,d,a,b)&&&x^{18}&=&(c,d,b,a)&&&x^{19}&=&(d,a,b,c)&&&x^{20}&=&(d,a,c,b)\\
x^{21}&=&(d,b,a,c)&&&x^{22}&=&(d,b,c,a)&&&x^{23}&=&(d,c,a,b)&&&x^{24}&=&(d,c,b,a)
\end{array}
$$
The following comparisons, whose validity can be readily 
established by the reader, demonstrate that for each $x^i$ in this set, 
$i=1,\ldots,24$, there is a distinct $x^j$ that 
intensity-dominates $x^i$---a situation denoted by $x^jDx^i$.
$$
\begin{array}{llllllll}
x^6Dx^1&x^1Dx^2&x^4Dx^3&x^{13}Dx^4&x^3Dx^5&x^{17}Dx^6&x^1Dx^7&x^2Dx^8\\
x^3Dx^9&x^8Dx^{10}&x^9Dx^{11}&x^{11}Dx^{12}&x^{18}Dx^{13}&x^{13}Dx^{14}
&x^{17}Dx^{15}&x^{22}Dx^{16}\\
x^{23}Dx^{17}&x^{24}Dx^{18}&x^8Dx^{19}&x^{22}Dx^{20}&x^{11}Dx^{23}&x^{12}Dx^{22}
&x^{24}Dx^{23}&x^{10}Dx^{24}
\end{array}
$$
Therefore, no intensity-efficient allocation exists in this profile.

The somewhat surprising fact that a plausible dominance concept 
may be cyclic and prevent an optimal entity to emerge invites an 
informal analogy between intensity-dominance cycles 
over \textit{allocations} with at least 4 agents and 
Condorcet cycles over \textit{alternatives} in pairwise 
majority-based preference aggregation with at least 3 agents
\citep{condorcet}. 
Importantly, however, although intensity-dominance cycles 
here may prevent refining the Pareto set, unlike the Condorcet 
social-welfare theoretic framework these do not lead to 
a ``policy paralysis'' problem because the Pareto set is always
non-empty and, absent any distributively juster suggestions, 
one of them might be promoted by the social planner. 
That said, it is naturally of interest to understand better the conditions
under which our proposed refinement of Pareto-efficiency is well-defined.
We turn to this problem next.

For any set $A'\subseteq A$ and preference relation $\succ_i$
that is induced by a strict intensity relation $\winteni$ 
we write $\succ_i^{A'}$ for the restriction of $\succ_i$ on $A'$.

\begin{dfn}
Given an intensity profile, 
the set $\{N'\subseteq N,A'\subseteq A\}$ is: 
\begin{enumerate}
\item \textnormal{Envy-free} 
if $|N'|=|A'|$, every agent $i\in N'$ has a 
distinct $\succ_i$-best item in $A$,
and this item (denoted $b_i'$) lies in $A'$.
\item \textnormal{Top-preference-coincident} 
if $\succ_i^{A'}=\succ_j^{A'}$ for all agents $i,j\in N'$ and, 
whenever $A'\subsetneq A$, it holds that $a'\succ_i a$ 
for all $a'\in A'$, $a\in A\setminus A'$, and all $i\in N'$.
\end{enumerate}
\end{dfn}
An envy-free set $\{N',A'\}$ comprises agents 
in $N'$ who have distinct overall-best items in $A'$. 
This terminology is motivated by the fact that,  
when $N'= N$ and $A'= A$, each agent receiving their 
most preferred option results in an \textit{envy-free}
allocation \citep{varian74,moulin04}.
A top-preference-coincident set $\{N',A'\}$
consists of agents whose preferences over 
$A'$ are identical and such that everything in $A'$ is 
preferred to everything in $A\setminus A'$ (their
preferences over the latter subset may differ).

\begin{dfn}
Given a profile, a top-preference-coincident set $\{N',A'\}$
with the common preference ordering 
$a_1\succ^{A'}a_2\succ^{A'}\cdots \succ^{A'}a_{|A'|}$ is:

\begin{enumerate}
\item \textnormal{Monotonic}  if $|N'|=|A'|$ and
there is some permutation $\pi$ on $N'$ such that

\vspace{5pt}

\begin{tabular}{ccccccccc}
$s_{\pi(1)}(a_{1},a_{2})$ 
& $>$ 
& $s_{\pi(1)}(a_{2},a_{3})$ 
& $>$ 
& $\cdots$ 
& $>$ 
& $s_{\pi(1)}(a_{|A'|-1},a_{|A'|})$
& $>$ 
& $0$ \\
$\geq$ 
& 
& $\geq$ 
& 
& $\cdots$ 
& 
& $\geq$ 
&
&\\
$s_{\pi(2)}(a_{1},a_{2})$ 
& $>$ 
& $s_{\pi(2)}(a_{2},a_{3})$ 
& $>$ 
& $\cdots$ 
& $>$ 
& $s_{\pi(2)}(a_{|A'|-1},a_{|A'|})$
& $>$ 
& $0$ \\
$\geq$ 
& 
& $\geq$ 
& 
& $\cdots$ 
& 
& $\geq$ 
&
& \\
$\vdots$ 
& 
& $\vdots$ 
& 
& $\vdots$ 
& 
& $\vdots$
&
& $\vdots$ \\
$\geq$ 
& 
& $\geq$ 
& 
& $\cdots$ 
& 
& $\geq$ 
& 
&\\
$s_{\pi(|N'|)}(a_{1},a_{2})$ 
& $>$ 
& $s_{\pi(|N'|)}(a_{2},a_{3})$ 
& $>$ 
& $\cdots$ 
& $>$ 
& $s_{\pi(|N'|)}(a_{|A'|-1},a_{|A'|})$
& $>$ 
& $0$
\end{tabular}

\item \textnormal{Strictly monotonic} if it is
monotonic and all inequalities above are strict. 

\item \textnormal{Homeo-monotonic} if it is monotonic 
and, for all $i,j$ such that $a_i\succ^{A'} a_j$,

\vspace{5pt}

\begin{tabular}{lll}
	$s_{\pi(i)}(a_i,a_j)$ & $\geq$ & $s_{\pi(j)}(a_i,a_j)$.
\end{tabular}
\end{enumerate}
\label{dfn:homeo-monotonic-set}
\end{dfn}

Suppose agents 1, 2 and 3 all put items $a$, $b$ and $c$
at the top of their preferences, and in that order
(that is, the permutation $\pi$ in the above definition is
such that $\pi(i)=i$ for $i=1,2,3$). 
Suppose further that the intensity difference 
between $a$ and $b$ is ranked higher 
for the first agent than for the second, 
which in turn is ranked higher than 
for the third agent (i.e. $s_1(a,b)\geq s_2(a,b)\geq s_3(a,b)$).
Suppose the same is true for the intensity difference between
$b$ and $c$. Then, the top-preference-coincident 
set $\{\{1,2,3\},\{a,b,c\}\}$ is monotonic in the sense of
Definition \ref{dfn:homeo-monotonic-set}:
there is a weak ordering over agents according to which 
the intensity differences between any two consecutive items
in the agents' common preference ordering are ranked.
By itself, however, this does not say anything about the relative
ranking, among agents,  of intensity differences between 
non-consecutively ordered items. 
Homeo-monotonicity imposes the requirement that these, too, 
are interpersonally ranked in alignment with the order that 
underlies the successive items. 
In this example, this means that agent 1 prefers $a$ to $c$ 
at least as much as agent 2, who does so at least as much as 
agent 3, i.e. $s_1(a,c)\geq s_2(a,c)\geq s_3(a,c)$.
Monotonic sets that satisfy this additional condition 
are intuitive, as they paint a clear picture 
on which agent's preferences there are relatively more intense; 
which ones are second most intense; and so on.
As we show shortly in Theorems \ref{thm:homeomonot-neces} 
and \ref{thm-homeomonot} and the accompanying examples, 
such alignment is in subtle ways crucial for the existence 
of intensity-efficient allocations at a profile.

We remark that a homeo-monotonic set is monotonic by definition, but not 
necessarily strictly monotonic. 
Although this would 
have been implied by it, we will \textit{not} be assuming 
\textit{strict} homeo-monotonicity whereby all inequalities
in that definition are required to be strict.
Instead, the remaining two existence results of this section assume 
certain dichotomies\footnote{Our use of this term is unrelated to 
how it is used in \cite{bogomolnaia-moulin04} and several other
studies, where preference dichotomy refers to every agent's partitioning 
of all items into ``acceptable'' and ``unacceptable'', 
with the agent being indifferent between items in either subset.} 
(in fact, \textit{poly}chotomies) in agents' 
preferences and intensities. 

More specifically, 
we say that an intensity profile \textit{can be partitioned into 
monotonic sets and an envy-free set}, the latter possibly being empty,  
if there are pairs $\{N^i\subseteq N,A^i\subseteq A\}$, $i\leq m$, such that   
the collections defined by $\{N^i\}_{i=1}^m$, $\{A^i\}_{i=1}^m$ 
partition $N$, $A$, and each $\{N^i,A^i\}$ is monotonic or envy-free. 
In other words, the sets of 
agents and goods can be partitioned into certain cells $\{N^i,A^i\}$ 
that are in a one-to-one correspondence with each other 
in the sense that the number of agents in $N^i$ is the same
as the number of items in the corresponding set $A^i$. 
These cells can be of the envy-free type 
(by definition, at most one such cell exists)
or of the monotonic type. 
Agents in the first cell have completely heterogeneous
preferences as far as their top items are concerned. 
By virtue of the top-preference-coincidence implication of 
monotonicity, agents within each cell of the second kind have identical 
preferences up to a certain rank-order position 
(say, the 3rd-ranked item for a cell with 3 elements), 
while agents across cells have distinct such preferences
at the top. Within each such cell, finally, 
agents can be weakly or strictly ordered in the way in which they rank 
their intensity of preference between any two items that appear 
consecutively in the relevant upper part of their
preference ordering.
Clearly, any set of agents can be partitioned into groups 
according to whether distinct or identical alternatives are 
the most preferred within each group.
The additional structure imposed here 
requires that: (i) the preferences of agents with 
the same top-ranked item also coincide in how
they rank their second-, third-, $\ldots$, $k$th-most preferred
item for some $k\leq n$ that coincides with the number of agents
in that group; (ii) the preferences of distinct 
such groups are sufficiently heterogeneous; 
and (iii) within each group, agents are ordered in the way 
in which they rank their intensity differences between 
any two items that appear successively in their preference order.

\begin{thm}[Necessary Condition]\label{thm:homeomonot-neces}
If a profile has an intensity-efficient allocation and 
can be partitioned into monotonic sets and an envy-free set, 
then every strictly monotonic set in it is homeo-monotonic.
\end{thm}

This result is proved by showing that 
if such a profile harbours a strictly monotonic but not homeo-monotonic 
set, every Pareto-efficient allocation is intensity-dominated.
Since intensity-efficient allocations are Pareto-efficient and 
not dominated, the claim follows. 

The next two examples, respectively, 
illustrate Theorem \ref{thm:homeomonot-neces} 
and explain why this result 
cannot be strengthened by replacing 
``strictly monotonic'' with ``monotonic'' in 
the consequent part of its statement.

\begin{example}\label{exm:counterexample5}
For $n=5$ let the envy-free
and top-preference-coincident sets 
that partition $N$ and $A$ be defined by 
$\{N^1,A^1\}=\{\{4,5\},\{a_4,a_5\}\}$ and 
$\{N^2,A^2\}=\{\{1,2,3\},\{a_1,a_2,a_3\}\}$,
as in Table \ref{tab:strictly-monotonic-not-homeomonotonic}.
	
\begin{table}[!htbp]
	\centering
	\arraycolsep=0pt\def\arraystretch{1.2}
	\small
	\caption{\centering
		A profile with a strictly monotonic but not 
		homeo-monotonic set 
		and no intensity-efficient allocation.}
	\begin{tabular}{c|ccccc}
		& \multicolumn{3}{c}{
			$\overbrace{\hspace{40pt}\text{$N^2$}\hspace{40pt}}$} 
		& \multicolumn{2}{c}{$\overbrace{\hspace{20pt}\text{$N^1$\hspace{20pt}}}$\hspace{10pt}}
		\\
		& 1 
		& 2 
		& 3 
		& \;\; 4 
		& \;\; 5
		\\
		\backslashbox{\small\hspace{-10pt}$s_i(a,b)=$}{\hspace{-10pt}\small$i=$}
		& \multicolumn{3}{c}{\footnotesize 
			\text{$a_1\succ_ia_2\succ_ia_3\succ_ia_4\succ_ia_5$}}
		& \multicolumn{2}{c}{\footnotesize 
			$a_4\succ_4 \cdots$ $|$ $a_5\succ_5 \cdots$} 
		\\
		\hline
		10		& $(a_1,a_5)$
		& $(a_1,a_5)$
		& $(a_1,a_5)$
		& $(a_4,\cdot)$ 
		& $(a_5,\cdot)$ \\
		9		& $(a_1,a_4)$
		& $(a_1,a_4)$
		& $(a_1,a_4)$
		& $\vdots$ & $\vdots$ \\
		8		& $(a_2,a_5)$
		& $(a_2,a_5)$
		& \color{red} $(a_1,a_3)$
		&&\\
		7		& $(a_2,a_4)$
		& \color{red} $(a_1,a_3)$
		& $(a_2,a_5)$
		&&\\
		6		& \color{red}$(a_1,a_3)$
		& $(a_2,a_4)$
		& $(a_2,a_4)$
		&&\\
		5		& $(a_3,a_5)$
		& $(a_3,a_5)$
		& $(a_3,a_5)$
		&&\\
		4		& \color{magenta}$(a_1,a_2)$
		& $(a_3,a_4)$
		& $(a_3,a_4)$
		&&\\
		3		& \color{blue}$(a_2,a_3)$
		& \color{magenta}$(a_1,a_2)$
		& $(a_4,a_5)$
		&&\\
		2		& $(a_3,a_4)$
		& \color{blue}$(a_2,a_3)$
		& \color{magenta}$(a_1,a_2)$
		&&\\
		1		& $(a_4,a_5)$
		& $(a_4,a_5)$
		& \color{blue}$(a_2,a_3)$
		&&
	\end{tabular}
	\label{tab:strictly-monotonic-not-homeomonotonic}
	\end{table}
	
	\noindent
	The Pareto-efficient allocations here are: 
	$$
	\hspace{-8pt}
	\begin{array}{lllllll}
		\underbrace{(a_3,a_1,a_2,a_4,a_5)}_{s}, & 
		\underbrace{(a_3,a_2,a_1,a_4,a_5)}_{t}, & 
		\underbrace{(a_1,a_2,a_3,a_4,a_5)}_{x}, & \\ 
		\underbrace{(a_1,a_3,a_2,a_4,a_5)}_{y}, & 
		\underbrace{(a_2,a_3,a_1,a_4,a_5)}_{w}, & 
		\underbrace{(a_2,a_1,a_3,a_4,a_5)}_{z}	
	\end{array}
	$$
	The intensity-dominance relation $D$ is cyclic over this set:
	$$\small 
	\begin{array}{lllll}
		\hspace{-10pt}\overbrace{(a_3,\color{red}a_1,a_2\color{black},a_4,a_5) D (a_3,\color{red}a_2,a_1,\color{black}a_4,a_5)}^{sDt:\;\; s_2(a_1,a_2)>s_3(a_1,a_2)}&&&&\\ 
		&\hspace{-50pt}\overbrace{(\color{red}a_3,\color{black}a_2,\color{red}a_1,\color{black}a_4,a_5)D(\color{red}a_1,\color{black}a_2,\color{red}a_3,\color{black}a_4,a_5)}^{tDx:\;\; s_3(a_1,a_3)>s_1(a_1,a_3)}&&&\\ 
		&&\hspace{-90pt}\overbrace{(a_1,\color{red}a_2,a_3\color{black},a_4,a_5)D(a_1,\color{red}a_3,a_2\color{black},a_4,a_5)}^{xDy:\;\; s_2(a_2,a_3)>s_3(a_2,a_3)}&&\\ 
		&&&\hspace{-130pt}\overbrace{(\color{red}a_1,\color{black}a_3,\color{red}a_2\color{black},a_4,a_5)D(\color{red}a_2,\color{black}a_3,\color{red}a_1\color{black},a_4,a_5)}^{yDw:\;\; s_1(a_1,a_2)>s_3(a_1,a_2)}&\\ 
		&&&&\hspace{-140pt}\overbrace{(a_2,\color{red}a_3,a_1\color{black},a_4,a_5)D(a_2,\color{red}a_1,a_3\color{black},a_4,a_5)}^{wDz:\;\; s_3(a_1,a_3)>s_2(a_1,a_3)} \\ 
		\multicolumn{5}{c}{\hspace{-20pt}\underbrace{(\color{red}a_2\color{black},a_1,\color{red}a_3\color{black},a_4,a_5)D\color{red}(a_3\color{black},a_1,\color{red}a_2,\color{black}a_4,a_5)}_{zDs:\;\; s_1(a_2,a_3)>s_3(a_2,a_3)}}
	\end{array}
	$$

	\hfill$\blacklozenge$
\end{example}

\begin{example}\label{exm:strict-monot}
	For $n=4$ let the envy-free 
	and top-preference-coincident sets 
	that partition $N$ and $A$ be defined by 
	$\{N^1,A^1\}=\{\{4\},\{a_4\}\}$ and 
	$\{N^2,A^2\}=\{\{1,2,3\},\{a_1,a_2,a_3\}\}$,
	as in Table \ref{tab:monotonotic-not-homeomonotonic}.
	The Pareto-efficient allocations are 
	$p:=(a_1,a_2,a_3,a_4)$, 
	$q:=(a_1,a_3,a_2,a_4)$, $r:=(a_2,a_3,a_1,a_4)$,
	$x:=(a_2,a_1,a_3,a_4)$, $y:=(a_3,a_1,a_2,a_4)$ and  
	$z:=(a_3,a_2,a_1,a_4)$. 
	Among them, 
	$r$ is intensity-efficient: 
	it dominates $x$; is
	incomparable to $p$ and $y$;
	and is equivalent to 
	$q$, $z$ in the sense defined below (both these allocations 
	are dominated by $p$, $y$, respectively). 
	Yet $N^2$ is monotonic (not strictly so) but 
	not homeo-monotonic with respect to $A^2$: 
	$s_1(a_1,a_2)=s_2(a_1,a_2)>s_3(a_1,a_2)$ 
	and $s_1(a_2,a_3)=s_2(a_2,a_3)>s_3(a_2,a_3)$, 
	but $s_1(a_1,a_3)=s_2(a_1,a_3)<s_3(a_1,a_3)$.
	\hfill$\blacklozenge$
\end{example}

\begin{table}[!htbp]
	\centering
	\arraycolsep=0pt\def\arraystretch{1.2}
	\small
	\caption{\centering 
		A profile with a monotonic but not homeo-monotonic
		set and an intensity-efficient allocation.}
	\begin{tabular}{c|cccc}
		& \multicolumn{3}{c}{
			$\overbrace{\hspace{40pt}\text{$N^2$}\hspace{40pt}}$} 
		& $\overbrace{\hspace{7pt}\text{$N^1$\hspace{7pt}}}$ 
		\\
		& 1 
		& 2 
		& 3 
		& 4 
		\\
		\backslashbox{\small\hspace{-10pt}$s_i(a,b)=$}{\hspace{-10pt}\small$i=$}
		& \multicolumn{3}{c}{\footnotesize 
			\text{$a_1\succ_ia_2\succ_ia_3\succ_ia_4$}}
		& \text{\footnotesize 
			$a_4\succ_4 \cdots \succ_4 a_1$} 
		\\
		\hline
		\multicolumn{1}{c|}{\small 6}
		& $(a_1,a_4)$ 
		& $(a_1,a_4)$ 
		& $(a_1,a_4)$ 
		& $(a_4,a_1)$ 
		\\
		\multicolumn{1}{c|}{\small 5}
		& $(a_2,a_4)$  
		& $(a_2,a_4)$  
		& \color{red}$(a_1,a_3)$ 
		& \vdots
		\\
		\multicolumn{1}{c|}{\small 4}
		& \color{red}$(a_1,a_3)$ 
		& \color{red}$(a_1,a_3)$
		& $(a_2,a_4)$ 
		& 
		\\
		\multicolumn{1}{c|}{\small 3}
		& \color{magenta}$(a_1,a_2)$
		& \color{magenta}$(a_1,a_2)$
		& $(a_3,a_4)$  
		& 
		\\
		\multicolumn{1}{c|}{\small 2}
		& \color{blue}$(a_2,a_3)$
		& \color{blue}$(a_2,a_3)$
		& \color{magenta}$(a_1,a_2)$
		& 
		\\
		\multicolumn{1}{c|}{\small 1}
		& $(a_3,a_4)$ 
		& $(a_3,a_4)$ 
		& \color{blue}$(a_2,a_3)$
		& 
		\\
	\end{tabular}
	\label{tab:monotonotic-not-homeomonotonic}
\end{table}

The next result modifies the partitioning structure of the 
assumption in Theorem \ref{thm:homeomonot-neces} by imposing 
homeo-monotonicity. More specifically, we say that 
an intensity profile \textit{can be partitioned into 
homeo-monotonic sets and an envy-free set}, 
the latter possibly being empty, 
if there are pairs $\{N^i\subseteq N,A^i\subseteq A\}$, $i\leq m$, 
such that the collections defined by 
$\{N^i\}_{i=1}^m$, $\{A^i\}_{i=1}^m$ 
partition $N$, $A$, and each $\{N^i,A^i\}$ is homeo-monotonic 
or envy-free. 
Compared to the partitioning structure of 
Theorem \ref{thm:homeomonot-neces}, the additional requirement 
here is that all monotonic cells are also homeo-monotonic. 
That is, the relevant ordering over agents within any monotonic cell 
is required now to also be preserved in the way in which 
intensity differences are ranked between non-consecutive items 
under that cell's preferences.

\begin{thm}[Sufficient Condition]\label{thm-homeomonot}
If a profile can be partitioned into homeo-monotonic sets 
and an envy-free set, then it has an 
intensity-efficient allocation.
\end{thm}

This condition is not necessary: one can easily construct 
profiles with intensity-efficient allocations where it is 
violated. 
Yet, in light of Theorem \ref{thm:homeomonot-neces}, it is evident 
that some form of a homeo-monotonic structure is important 
for the existence of such allocations when 
subsets of agents and items can be clustered into 
monotonic or strictly monotonic sets, or variations thereof. 
The proof of Theorem \ref{thm-homeomonot}, below, 
shows how this specific structure is helpful toward 
clarifying some important mechanisms for the existence 
of intensity-efficient allocations in general. 
We return to these points after Example \ref{exm:illustration}.\\ 

\noindent \textbf{\textit{Proof of Theorem \ref{thm-homeomonot}.}}
Consider a profile $S$ with the postulated structure.
Assuming it is non-empty, 
let $N^1:=\{n^1_1,\ldots,n^1_{|N^1|}\}$ be the envy-free
set with respect to $A^1:=\{a^1_1,\ldots,a^1_{|A^1|}\}$,
and let $N^{i>1}:=\{i_1,\ldots,i_{|N^i|}\}$ be 
the $i$th monotonic set with respect to 
$A^{i>1}:=\{i_1,\ldots,i_{|A^i|}\}$, $i=2,\ldots,m$.
By the profile assumption, $|A^i|=|N^i|$ for all $i\leq m$,   
and the collections $(N^i)_{i=1}^m$ and $(A^i)_{i=1}^m$ 
partition the sets $N$ and $A$, respectively.
Without loss of generality we may assume that $a^1_j$ is the top-ranked
item of agent $n^1_j$, $j=1,\ldots,|A^1|$
(for the general case we apply a permutation over 
$\{1,\ldots,|A^1|\}$ on the $a^1$-subscripts).
Recalling our earlier notation, 
this amounts to writing $a_1^1\equiv b_1^1$, $a_2^1\equiv b_2^1$, etc.
Also without loss, for $A^{i>1}$, $N^{i>1}$ we may assume that
$a^i_1\succ_j a^i_2 \succ_j \cdots \succ_j a^i_{|A^i|}$ for all 
$j\in N^{i>1}$. 
(Note: the special cases where $N^1=\emptyset$, $N^1=N$ 
or $N^i=N$ for some $i>1$ are allowed in the postulated 
profile, and so will they be in the proof.)

\begin{lem}\label{lem:pareto-necessary}
	If a profile can be partitioned into an envy-free set
	$\{N^1,A^1\}$ (which may be empty) and monotonic sets
	$\{N^i,A^i\}$, $i>2$, then an allocation $x$ 
	is Pareto-efficient with respect to 
	$P_S$ if and only if it satisfies the following:\\
	(i) $j\in N^1$ $\Rightarrow$ $x_j=b_j^1$.\\
	(ii) $j\in N^{i>1}$ $\Rightarrow$ $x_j\in A^{i>1}$.\\
\end{lem}

\vspace{-20pt}

\begin{proof}
	For the ``if'' part, let $x$ be an arbitrary allocation
	that satisfies (i) and (ii). Assume to the contrary that $x$
	is not Pareto efficient. 
	Let $y$ be another allocation that Pareto-dominates $x$.
	This implies $s_l(y_l,x_l)\geq 0$ for every $l\leq n$, 
	with strict inequality for some $l$. 
	Let $N^*\subsetneq N$ consist of every agent $l$ 
	for whom $s_l(y_l,x_l)=0$.
	By Strictness, $s_l(y_l,x_l)=0 \Leftrightarrow y_l=x_l$.
	Suppose $N^*=\emptyset$. 
	From the above, this implies $s_l(y_l,x_l)> 0$ for all $l\leq n$,
	with every agent assigned a different item under $x$ and $y$.
	If the envy-free set $N^1$ is 
	non-empty, then this readily contradicts (i) for every 
	$l\in N^1$.
	If $N^1=\emptyset$, then there is at least one $i>1$ such that 
	the (monotonic in $A^{i>1}$) set $N^{i>1}$ is non-empty.
	Since $|N|=|A|=n$ holds by assumption, and recalling the notation 
	that precedes the statement of Lemma \ref{lem:pareto-necessary}, 
	it follows from (ii) that for the specific agent 
	$i_1\in N^{i>1}$ we must have $s_1^i(x_1^i,x'_i)>0$ 
	for all $x'_i\in A\setminus\{x_1^i\}$, a contradiction.
	Now suppose $N^*\neq\emptyset$. Since $y$ Pareto-dominates $x$, 
	the set $N^{**}:=N\setminus N^*$ that comprises every agent $l$ for 
	whom $s_l(y_l,x_l)>0$ holds is non-empty.
	From the preceding argument we know that for every $l\in N^{**}$ 
	it must be that $l\not\in N^1$ and $l\neq i_1\in N^{i>1}$, 
	for all $i>1$.
	By $s_l(y_l,x_l)>0$ and $|A|=|N|$, there is some agent $k\neq l$ such that 
	$y_l=x_k$. By Strictness and Pareto-dominance, $s_k(y_k,x_k)>0$.
	Since both $k$ and $l$ are better off at $y$, it follows that 
	$l\in N^{i>1}$ and $k\in N^{j>1}$ for distinct 
	sets $N^{i>1}$ and $N^{j>1}$ that are monotonic in 
	$A^{i>1}$ and $A^{j>1}$, respectively. 
	By the partition assumption, $A^i\cap A^j=\emptyset$.
	Since, by the definition of monotonic sets, 
	$s_l(a,a')>0$ and $s_k(b,b')>0$ holds 
	for all $a\in A^{i>1},a'\in A\setminus A^{i>1}$ 
	and $b\in A^{j>1},b'\in A\setminus A^{j>1}$,
	and since, by (ii), it holds that $x_l\in A^{i>1}$ 
	and $x_k\equiv y_l\in A^{j>1}$, the postulate 
	$s_l(y_l,x_l)>0$ leads to a contradiction.
	Therefore, $x$ is Pareto-efficient.

	For the ``only if'' part, consider an arbitrary 
	Pareto-efficient allocation $x$.
	Suppose to the contrary that (i) is false.
	That is, $x_j\neq b_j^1$ for some $j\in N^1$. 
	Let $b_j^1=x_l$ for the relevant distinct agent $l\in N$ 
	(such an agent exists because $|N|=|A|$). 
	Either $l\in N^1$ or $l\in N^{i>1}$.
	In the former case, swapping $x_j$ and $x_l$ between the two agents
	Pareto-improves upon $x$,
	contradicting its postulated Pareto optimality.
	In the latter case, from the assumption that $N^{i>1}$ 
	is monotonic in $A^{i>1}$, and by the partition assumption, we have 
	$a^i_t\succ_l x_l\equiv b^1_j$ for every $a^i_t\in A^{i>1}$.
	Furthermore, in this case (again because $|N|=|A|$)
	there must also exist another agent 
	$k\in N^{p}$, $p\neq i$, such that $x_k\in A^{i>1}$. 
	There are two possibilities: $p=1$ or $p>1$.
	If $p=1$, then the above contradiction is obtained now for 
	$x_l$ and $x_k$.
	If $p>1$, then, because $\{N^{l>1},A^{l>1}\}$ and  
	$\{N^{p>1},A^{p>1}\}$ are distinct cells of the partition, 
	we know that $s_l(x_k,x_l)>0$ and $s_k(x_l,x_k)>0$,
	which also contradicts the postulated Pareto-efficiency of $x$.		
	The argument establishing necessity of (ii) is analogous
	and omitted. 
\end{proof}

Next, consider a monotonic set $N^{i>1}$ with respect to $A^{i>1}$. 
By the stated assumption,
either such a non-empty set exists or the envy-free
set $N^1$ coincides with $N$. 
In the latter case, the allocation $x^*$ defined by 
$x^*_i=b^1_i$ is obviously the unique Pareto-efficient allocation 
at this profile and therefore, trivially, intensity-efficient.
So, in the sequel we assume that $N^{i>1}\neq\emptyset$.

Denote by 
$a_1^i\succ_i a_2^i \succ_i \cdots \succ_i a_{|A^{i}|}^i$  
the induced preference order on $A^{i>1}$ which, 
by the definition of $\{N^{i>1},A^{i>1}\}$, 
is common across all agents in $N^{i>1}$.
Define the binary relation $\trianglerighteq^i$ on $N^{i>1}$ by 
\begin{eqnarray*}
	v_j \trianglerighteq^i v_k & \Longleftrightarrow & 
	s_j(a^i_l,a^i_{l+1})\geq s_k(a^i_l,a^i_{l+1}) 
	\text{ for all } l=1,\ldots,|A^{i>1}|-1
\end{eqnarray*}
By the postulated monotonicity of $\{N^{i>1},A^{i>1}\}$, 
$\trianglerighteq^i$
is a weak order on $N^{i>1}$.
Let $\trianglerighteq_i^*$ be some extension of 
$\trianglerighteq^i$ into a linear order over $N^{i>1}$,
and denote by $\trianglerighteq^{*i}(j)$ the 
$\trianglerighteq^{*i}$-rank of agent $j\in N^{i>1}$.
By construction, $\trianglerighteq^{*i}$ coincides with $\trianglerighteq^{i}$ 
except where it breaks any equivalence ties that the latter 
relation may feature.

Now define allocation $x^*$ by
\begin{eqnarray*}
	x_j^{*1} := b_j^1,\; j=1,\ldots,|N^1| \qquad & \Longleftrightarrow & 
	\qquad j\in N^1\\
	x_j^{*i>1} := a^i_l,\; j=1,\ldots,|N^{i>1}| \qquad & \Longleftrightarrow 
	& \qquad j\in N^{i>1} \text{ and } \trianglerighteq^{*i}(j)=l
\end{eqnarray*}
That is, each agent in the set $N^1$ (which could be empty) 
is assigned their most preferred item,
while the $\trianglerighteq^{*i}$-$l$th-ranked agent in $N^{i>1}$
is assigned their $l$th most preferred item. 
Recalling that $N^1$ is 
envy-free and each $N^{i>1}$ is monotonic with respect to 
$A^{i>1}$,
it follows from Lemma \ref{lem:pareto-necessary} that 
$x^*$ is Pareto-efficient. 
We will prove that it is also intensity-efficient.

To this end, suppose to the contrary that some Pareto-efficient
allocation $y$ intensity-dominates $x^*$.
Then, there are agents $i,j\in N$ such that $(x_i^*,x_j^*)=(y_j,y_i)$
and $s_j(y_j,y_i)\equiv s_j(x_i^*,x_j^*)>s_i(x^*_i,x_j^*)>0$. 
By construction of $x^*$, $i,j$ cannot belong to the same 
$N^{l\geq 1}$. 
For $l=1$ this is obvious because $x^{*i}=b^1_i$ and $x^{*j}=b^1_j$, 
contradicting the above postulate.
For $l>1$, moreover, one has $x^*_i,x_j^*\in A^{l>1}$ 
while items in $A^{l>1}$ are assigned 
to agents in $N^{l>1}$ according to 
the linear order $\trianglerighteq^{*l}$, which, by 
homeo-monotonicity of $\{N^{l>1},A^{l>1}\}$, ensures 
that $s_i(x_i^*,x_j^*)\geq s_j(x_i^*,x_j^*)$, thereby 
contradicting that postulate.
The remaining possibility is 
$i\in N^{l>1}$ and $j\in N^{k>1}$ 
for some $k\neq l$. 
By construction of $x^*$, however, we have $x_i^*\in A^{l>1}$ and 
$x_j^*\in A^{k>1}$ which---because $N^{l>1}$, $N^{k>1}$ are 
monotonic with respect to $A^{l>1}$, $A^{k>1}$,
and $A^{l>1}\cap A^{k>1}=\emptyset$ also holds by the
partitioning assumption---implies 
$s_i(x_i^*,x_j^*)\equiv s_i(y_j,y_i)>0$
and $s_j(x_i^*,x_j^*)\equiv s_j(y_j,y_i)<0$. 
Thus, the postulate that $y$ intensity-dominates $x^*$ 
is contradicted in all situations.
\hfill $\blacksquare$\\

The sufficient condition of Theorem \ref{thm-homeomonot} is trivially
satisfied if $\{N,A\}$ is itself envy-free.
The example that follows shows how it can be satisfied non-trivially, 
and also illustrates one additional relevant 
concept that we now introduce.

\begin{dfn}
	Two allocations $x$ and $y$ that are 
	intensity-efficient at a profile 
	are \textnormal{equivalent} if $s_i(x_i,x_j)=s_j(y_j,y_i)$ 
	for all pairs of agents $(i,j)$ such that $(x_i,x_j)=(y_j,y_i)$, 
	and $x_k=y_k$ for every other agent $k\leq n$.
\end{dfn}

\noindent 
That is, two intensity-efficient allocations are 
equivalent if the agents in every ``flipping'' pair 
where the same items are assigned but in opposite ways have the same 
preference intensity for their preferred over their dis-preferred one,  
in the sense that the intensity difference between these items 
is ranked the same way in the two agents' intensity orderings.

\begin{example}\label{exm:illustration}
	Table \ref{exm:effect-homeomonotonic} describes a 
	profile %$S\in\mathcal{\widehat{S}}$ 
	where $n=6$ and 
	the envy-free and top-preference-coincident sets 
	that partition $N$, $A$ are
	$\{N^1,A^1\}=\{\{1\},\{a_1\}\}$, 
	$\{N^2,A^2\}=\{\{2,3,4\}$, $\{a_2,a_3,a_4\}\}$
	and  
	$\{N^3,A^3\}=\{\{5,6\},\{a_5,a_6\}\}$.

\begin{table}[!htbp]
	\centering
	\arraycolsep=0pt\def\arraystretch{1.2}
	\small
	\caption{\centering 
	A profile that can be partitioned 
	into an envy-free and two homeo-monotonic sets.}
	\begin{tabular}{c|cccccc}
		& $\overbrace{\hspace{7pt}\text{$N^1$\hspace{7pt}}}$ 
		& \multicolumn{3}{c}{
			$\overbrace{\hspace{40pt}\text{$N^2$}\hspace{40pt}}$} 
		& \multicolumn{2}{c}{
			$\overbrace{\hspace{20pt}\text{$N^3$}\hspace{20pt}}$}
		\\
		& \color{magenta}1 
		& \color{red}2 
		& \color{red}3 
		& \color{red}4 
		& \color{blue}5 
		& \color{blue}6
		\\
		\backslashbox{\small\hspace{-8pt}$s_i(a,b)=$}{\hspace{-8pt}\small$i=$}
		& \text{\footnotesize 
			${\color{magenta}a_1}\succ_1\cdots$} 
		& \multicolumn{3}{c}{\footnotesize 
			\text{${\color{red}a_2\succ_ia_3\succ_ia_4}\succ_i\cdots$}}
		& \multicolumn{2}{c}{\footnotesize 
			\text{${\color{blue}a_5\succ_ia_6}\succ_i\cdots$}}\\
		\hline
		\multicolumn{1}{c|}{\small 15}
		&\color{magenta}$(a_1,a_6)$ 
		&$(a_2,a_1)$ 
		&$(a_2,a_5)$ 
		&$(a_2,a_6)$ 
		&$(a_5,a_1)$
		&$(a_5,a_3)$
		\\
		\multicolumn{1}{c|}{\small 14}
		&$(a_2,a_6)$
		&$(a_2,a_6)$  
		&$(a_3,a_5)$ 
		&$(a_3,a_6)$ 
		&$(a_5,a_4)$
		&$(a_5,a_1)$
		\\
		\multicolumn{1}{c|}{\small 13}
		&$(a_1,a_5)$
		&$(a_2,a_5)$  
		&$(a_2,a_6)$ 
		&$(a_4,a_6)$ 
		&$(a_5,a_3)$ 
		&$(a_5,a_4)$
		\\
		\multicolumn{1}{c|}{\small 12}
		&$(a_3,a_6)$
		&\color{red}$(a_2,a_4)$
		&$(a_3,a_6)$  
		&$(a_2,a_1)$  
		&$(a_5,a_2)$ 
		&$(a_5,a_2)$
		\\
		\multicolumn{1}{c|}{\small 11}
		&$(a_2,a_5)$
		&$(a_3,a_1)$ 
		&$(a_2,a_1)$  
		&$(a_3,a_1)$ 
		&$(a_6,a_1)$ 
		&$(a_6,a_3)$
		\\
		\multicolumn{1}{c|}{\small 10}
		&$(a_1,a_4)$
		&$(a_3,a_6)$ 
		&$(a_3,a_1)$ 
		&$(a_4,a_1)$ 
		&$(a_6,a_4)$ 
		&$(a_6,a_1)$
		\\
		\multicolumn{1}{c|}{\small 9}
		&$(a_4,a_6)$
		&$(a_3,a_5)$ 
		&$(a_4,a_5)$ 
		&$(a_5,a_6)$ 
		&\color{blue}$(a_5,a_6)$ 
		&\color{blue}$(a_5,a_6)$
		\\
		\multicolumn{1}{c|}{\small 8}
		&$(a_3,a_5)$
		&\color{red}$(a_2,a_3)$
		&\color{red}$(a_2,a_4)$ 
		&$(a_2,a_5)$ 
		&$(a_6,a_3)$
		&$(a_6,a_4)$
		\\
		\multicolumn{1}{c|}{\small 7}
		&$(a_2,a_4)$
		&\color{red}$(a_3,a_4)$
		&\color{red}$(a_2,a_3)$
		&$(a_3,a_5)$ 
		&$(a_2,a_1)$
		&$(a_2,a_3)$
		\\		
		\multicolumn{1}{c|}{\small 6}
		&$(a_1,a_3)$ 
		&$(a_4,a_1)$
		&\color{red}$(a_3,a_4)$ 
		&$(a_4,a_5)$
		&$(a_2,a_4)$
		&$(a_4,a_3)$\\
		\multicolumn{1}{c|}{\small 5}
		&$(a_5,a_6)$ 
		&$(a_4,a_6)$
		&$(a_4,a_6)$
		&$(a_5,a_1)$
		&$(a_3,a_1)$
		&$(a_2,a_1)$\\
		\multicolumn{1}{c|}{\small 4}
		&$(a_4,a_5)$ 
		&$(a_5,a_1)$
		&$(a_1,a_5)$
		&$(a_1,a_6)$ 
		&$(a_6,a_2)$
		&$(a_6,a_2)$\\
		\multicolumn{1}{c|}{\small 3}
		&$(a_3,a_4)$ 
		&$(a_4,a_5)$
		&$(a_4,a_1)$
		&\color{red}$(a_2,a_4)$ 
		&$(a_2,a_3)$
		&$(a_2,a_4)$\\
		\multicolumn{1}{c|}{\small 2}
		&$(a_2,a_3)$ 
		&$(a_5,a_6)$
		&$(a_1,a_6)$
		&\color{red}$(a_2,a_3)$ 
		&$(a_3,a_4)$
		&$(a_4,a_1)$\\
		\multicolumn{1}{c|}{\small 1}
		&$(a_1,a_2)$ 
		&$(a_6,a_1)$
		&$(a_6,a_5)$
		&\color{red}$(a_3,a_4)$
		&$(a_4,a_1)$
		&$(a_1,a_3)$	
	\end{tabular}
	\label{exm:effect-homeomonotonic}
\end{table}	
\noindent
By Lemma \ref{lem:pareto-necessary}, 
there are 12 Pareto-efficient allocations 
in this profile, namely those which assign 
$a_1$ to agent $1$ and alternate items 
$a_2,a_3,a_4$ among agents $2,3,4$ and 
$a_5,a_6$ among $5,6$. 
Out of these 12 allocations, only  
the following 2 are the (in fact, equivalent)
intensity-efficient ones:
$(a_1,a_2,a_3,a_4,a_5,a_6)$; 
$(a_1,a_2,a_3,a_4,a_6,a_5)$.
\hfill $\blacklozenge$
\end{example}

Further, we note that the way in which the particular intensity-efficient allocation 
(labelled $x^*$) is constructed in the proof of Theorem \ref{thm-homeomonot}
suggests the following stronger implication
for a class of special cases:

\begin{cor}
An intensity-dominant---up to equivalence---allocation exists 
in any profile that satisfies the condition of Theorem \ref{thm-homeomonot}
and is also such that every monotonic set has two agents.
\end{cor}

Theorems \ref{thm:homeomonot-neces}-\ref{thm-homeomonot}
and Examples \ref{exm:counterexample5}-\ref{exm:illustration}
highlight the relevance of homeo-monotonicity 
for the existence of intensity-efficient allocations
when agents' preferences and intensities are polychotomous 
and neatly structured within each cluster where agents' 
preferences coincide.
As the non-existence example 
in the proof of Theorem \ref{thm:n=3} shows, moreover, 
a profile may fail to have an intensity-efficient allocation 
when a subtle variation of homeo-monotonicity is violated. 
More specifically, that profile 
features: (i) $n=4$; (ii) $a \succ_i b \succ_i c \succ_i d$ for $i\leq 4$; 
(iii) $s_1=s_4$, $s_2=s_3$. 
Although the four agents' intensities associated with  
their common preferences do not define a \textit{monotonic} 
top-preference-coincident set over the 4 items, 
one observes that emergent here is a variant 
of this notion whereby each set in the partition 
$\{\{1,2\},\{3,4\}\}$
of $N$ is monotonically top-preference-coincident with respect 
to the subset $\{a,b,c\}\subset A$:
$s_1(a,b)>s_2(a,b)$, $s_1(b,c)=s_2(b,c)$ and 
$s_4(a,b)>s_3(a,b)$, $s_4(b,c)=s_3(b,c)$.
However, both these sets of agents violate what homeo-monotonicity 
would have required here, because
$s_1(a,c)<s_2(a,c)$ and $s_4(a,c)<s_3(a,c)$. 
Although the problem of fully characterizing the conditions under 
which strict profiles have intensity-efficient allocations remains open, 
in light of the preceding analysis we conjecture 
that a necessary and sufficient condition revolves around 
the absence of any set of agents and items---possibly of different
sizes---that are in some sense monotonic but not homeo-monotonic.

Finally, we remark that the paper's new analytical environment  
necessitates novel arguments in the proofs of this section's main 
results. 
More specifically, viewed as an abstract mathematical problem, existence of 
intensity-efficient allocations amounts to identifying sufficient conditions 
for acyclicity of the intensity-dominance binary relation defined 
on the set of permutations over alternatives. 
The mathematical literature on the combinatorics of permutations 
has studied partial orders over such sets, which may even have the
richer structure of a lattice that additionally ensures 
the existence of both a greatest (hence maximal) and a smallest element.
These results, however, are not applicable in our environment.
Indeed, the \textit{strong} and 
\textit{weak Bruhat} partial orders in question 
\citep[pp. 399-400]{stanley12}, 
relate two permutations if and only if there is a 
\textit{single} transposition (``flip'') between elements 
in the two permutations, and this transposition corresponds 
to an \textit{inversion} 
\citep[p. 30]{stanley12} relative to some postulated 
fixed linear order over its elements. 
The weak Bruhat order 
(see \citeauthor{ceballos-pons24}, \citeyear{ceballos-pons24}
for a recent generalization)
gives rise to a lattice, whose graph is known 
as the \textit{permutahedron} or \textit{permutohedron}, 
but under the 
additional assumption that the said single transposition occurs 
between \textit{adjacent} elements of the two relevant permutations.
Our problem, by contrast, does not assume a fixed linear order 
over items, and is defined by intensity-dominance comparisons 
that generally feature possibly \textit{multiple} 
transpositions of \textit{non-adjacent} elements. 
The latter, in turn, are defined by dominance between specific 
entries in the two relevant columns of an integer matrix, 
themselves dictated by the corresponding agents' intensity orderings.

\subsection{Analysis of Enumerated and Simulated Profiles}
\label{subsec:sims}

We now use exhaustive enumerations and simulations to 
assess how intensity-efficient 
allocations refine Pareto-efficient ones, and compare these
refinements to those achieved under Borda allocations 
(the latter are defined as those that maximize the total 
Borda scores [see \eqref{BO-0}] across agents).
In particular, we sampled randomly 1 million strict ordinal 
intensity profiles in each of the two cases where $n=4,5$,
and analysed the full set of distinct such profiles when $n=3$. 
We clarify that, as per the novel enumerations output shown in 
Table \ref{tab:enum}, there are $12^3=1728$, $384^4$ and $92,160^5$ 
such profiles. Our global enumeration for $n=3$ and 
simulations for $n=4,5$ use this enumerations output 
and draw from the full domain in each of the latter two cases.

\begin{table}[!htbp]
	\centering
	\footnotesize
	\caption{The number of distinct intensity relations of different
		classes for small values of $n$.}
	\setlength{\tabcolsep}{3pt} % Default value: 6pt
	\renewcommand{\arraystretch}{1.2} % Default value: 1
	\begin{tabular}{|c|c|c|c|c|}
		\hline 
		&&\textbf{Strict \& utility-difference}
		&\textbf{Strict}
		& \textbf{Strict ordinal}\\
		&\textbf{Strict}
		&\textbf{representable}
		&\textbf{ordinally representable}
		&\textbf{intensity} \\
		{ }$n$ { }
		&\textbf{orders}
		&\textbf{intensity relations}
		&\textbf{intensity relations}
		&\textbf{profiles}\\
		\hline
		3 & 6	& 12  		 & 12 		& $12^3$ \\
		4 & 24	& 240		 & 384 		& $384^4$  \\
		5 & 120	& 13,680 	 & 92,160	& $92,160^5$ \\
		\hline
	\end{tabular}	
	\label{tab:enum}
	
	\scriptsize 
	Source: author's computations 
	with the Minizinc constraint-programming tool 
	\citep{minizinc}.
\end{table}

\begin{table}[!htbp]
	\centering
	\small
	\caption{\centering
		Pareto-efficient, intensity-efficient and Borda allocations
		at all intensity profiles when $n=3$,  
		and at one million profiles drawn randomly from the full domain 
		when $n=4,5$.}
	\setlength{\tabcolsep}{3pt} % Default value: 6pt
	\renewcommand{\arraystretch}{1.2} % Default value: 1
	\makebox[\linewidth][c]{
		\begin{tabular}{|l|c|c|c|}
			\hline
			&$n=3$&$n=4$&$n=5$\\
			\hline
			Proportion of profiles where intensity-efficient allocations refine Pareto-efficient strictly 	
			&$\frac{2}{3}$ &88.8\%&95.7\% \\
			\hline
			Proportion of profiles where Borda allocations refine Pareto-efficient strictly 		
			&$\frac{1}{2}$ &81.8\%&93.1\% \\
			\hline
			Proportion of profiles where the sets of Borda \& intensity-efficient allocations coincide
			&$\frac{2}{3}$ 
			&35.7\%
			&16.8\% \\
			\hline
			Proportion of profiles where the sets of Borda \& intensity-efficient allocations are disjoint
			&0
			&8.6\%
			&23.4\% \\
			\hline
			Average number of Pareto-efficient allocations
			&2.72   &5.51  &12.24 \\
			\hline
			Average number of intensity-efficient allocations
			&1.55    &1.79  &2.39 \\
			\hline
			Average number of Borda allocations			
			&1.72 	&1.8   &1.93 \\
			\hline
		\end{tabular}
	}
	\label{tab:simulations}
\end{table}

\begin{figure}[!htbp]
	\centering
	\caption{\centering The distribution of intensity-efficient 
		allocations as a proportion of Pareto-efficient allocations within a given profile.}
	\includegraphics[width=0.9\linewidth]{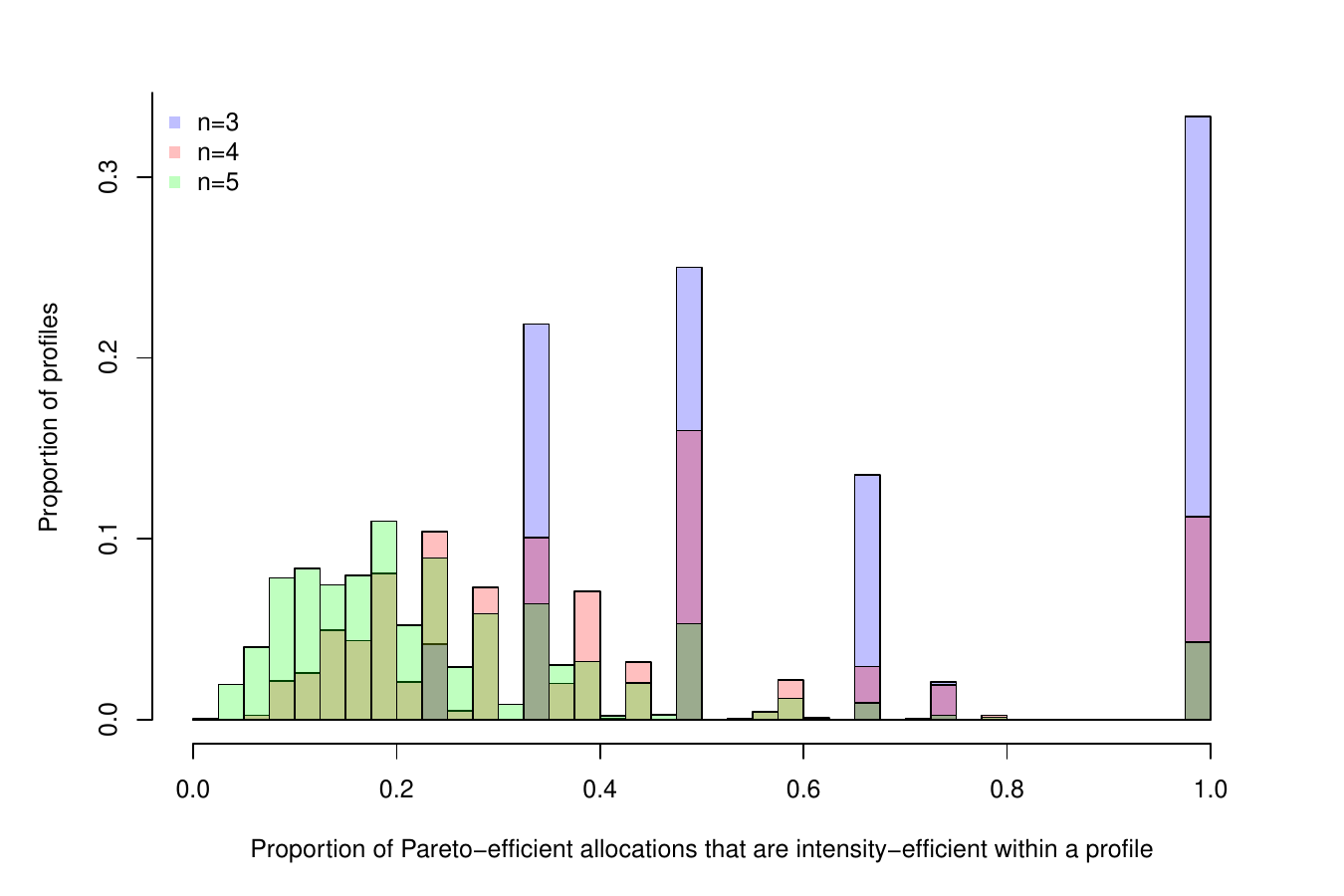}
	\label{fig:histograms}
\end{figure}

Table \ref{tab:simulations} summarizes the key findings 
of this exercise. The first two lines show the proportions 
of profiles where intensity-efficient and Borda 
allocations strictly refine the corresponding set of 
Pareto-efficient allocations. 
These proportions increase in $n$ and are uniformly higher
under intensity-efficiency. 
The next two lines further show that the sets of intensity-efficient 
and Borda allocations grow apart as $n$ increases. 
The last three lines, finally, show that the refinement gains 
under both criteria also increase in $n$, with the intensity-efficiency 
one discarding, on average,  approximately 43\%, 67.5\% and 80.5\% 
Pareto-efficient allocations as intensity-dominated when $n=3$, 4 and 5, respectively.
The refinement gains afforded by this criterion are therefore 
significant and, unlike those under Borda, 
are directly informed by agents' preference intensities,  
rather than their preferences used as proxies for their intensities.
Finally, non-existence of intensity-efficient allocations is a rare 
occurrence in these global simulations, 
with a $2\times 10^{-6}$ relative frequency when $n=4$ and 
$1.84\times 10^{-4}$ ($<0.0002)$ when $n=5$, respectively 
(these rare zeros were included in the computation of the statistics
of Table \ref{tab:simulations}).
Figure \ref{fig:histograms} supplements the above analysis with 
a graphical visualization that shows how the per-profile 
proportion of intensity-efficient allocations as a fraction 
of Pareto-efficient ones is distributed for the three different 
values of $n$ in this enumeration and simulation output. 

\section{Elicitation and Matching}

In this section we illustrate the potential usefulness of the preceding
analysis in matching-theoretic problems. 
In particular, we study a matching rule/mechanism 
that takes agents' ordinal intensities alongside their ordinal preferences
as its input and produces an allocation that improves upon Pareto-efficiency
in the intensity-dominating direction. 
Before doing so, however, we explain how ordinal intensities 
can be elicited alongside preferences in the first place 
in order for such a mechanism to become operational.
In particular, we show how strict intensity 
relations that are representable by ordinal preference 
intensity functions can simplify
this elicitation problem considerably.

To this end, observe first that the set of pairs 
of items in $A\times A$ contains $2{n\choose 2}+n$ elements. 
By Strictness, ${n\choose 2}$ of these pairs 
comprise distinct alternatives that are ranked by strict preference.
Moreover, also by Strictness, these pairs are themselves ranked 
strictly by any relation $\wintenl$ in the class under study. 
Therefore, a total of ${{n\choose 2}\choose 2}$ 
binary comparisons are involved in the $\wintenl$-ordering over 
this ``upper'' subset of pairs in $A\times A$. 
By \eqref{PIF2}, finally, (that is, by the generalized form of 
the skew-symmetry property in \eqref{canonical} that is built in 
the canonical intensity function $s_l$ which, by Lemma 
\ref{lem:canonical}, can always be chosen to represent $\wintenl$), 
the $\wintenl$-ordering over these pairs 
automatically determines the order over the full domain $A\times A$.
However, rather than requiring ${{n\choose 2}\choose 2}$ queries
($O(n^4)$), the algorithm that we construct below shows how 
the information in the underlying preference ordering, 
$\succ_l$, together 
with properties \eqref{PIF2} and \eqref{PIF3} (lateral 
consistency) of preference intensity 
functions, can be leveraged towards eliciting 
$\wintenl$ in the quadratically smaller number 
of ${n\choose 2}$ ($O(n^2)$) steps, 
and with ${n\choose 2}-2$ queries 
(that is two of the steps in this process do not 
require any new information).
As the algorithm's construction below shows, 
the first block of queries are binary, while 
the second block feature decisions from 
$n-1$, $n-2$, $\ldots$ 2 pairs of consecutively-ranked items.
The algorithm's decision tree is shown in 
Figure \ref{fig:elicit}. 

\begin{prp}
\label{prp:elicit}
Given a strict preference order,
a strict intensity relation 
that induces it can be elicited in $\binom{n}{2}$ steps---hence 
in $O(n^2)$ time---and with $\binom{n}{2}-2$ queries.
\end{prp}

\noindent \textbf{\textit{Proof of Proposition \ref{prp:elicit}}.}
Suppose that given is the strict preference
relation $\succ_l$ on $A$ that is defined by 
$a_1\succ_l a_2 \succ_l \ldots \succ_l a_n$.
Let $P:=\{(i,j):1\leq i<j\leq n\}$ and 
define the binary relation $\widehat{\succsim}_l$ on 
$Q:=\{(a_i,a_j)\in A\times A: (i,j)\in P\}$ by
\begin{eqnarray}
\label{elicit-eq1}
(a_i,a_j)\,\widehat{\succsim}_l\,(a_k,a_l) & 
\Longleftrightarrow & i\leq k \; \& \; j\geq l%\\
%\nonumber
%& \Longleftrightarrow & [i,j]\supseteq [k,l]
\end{eqnarray}
By definition, $\widehat{\succsim}_l$ is a partial order 
%(by interval containment) 
that aligns 
with $\succ_l$. We will extend $\widehat{\succsim}_l$ to a 
linear order $\wintenl$ on $Q$---hence, by \eqref{PIF2}, 
on $A\times A$ too---which induces $\succ_l$ on $A$.

To extend $\widehat{\succsim}_l$ to $\wintenl$ 
we must perform ${n\choose 2}$ rank placements, one 
for each pair $(a_i,a_j)$ with $(i,j)\in P$.
We write an algorithm that does so after $\binom{n}{2}$ steps
and $\binom{n}{2}-2$ queries that are raised 
in all but the first and last steps.
Each step of the algorithm involves selecting 
one of the $\widehat{\succsim}_l$-maximal pairs
that remain unranked. Although in the early steps of the 
process such selection is by means of binary queries, 
in later steps a query can involve selecting from 
up to $n-1$ pairs. By construction of $\widehat{\succsim}_l$, 
this maximum size corresponds to a maximal set comprising 
the pairs of items that are ranked consecutively by $\succ_l$, i.e. 
$(a_1,a_2)$, $(a_2,a_3)$, $\ldots$, $(a_{n-1},a_n)$.\\

\noindent 
\textit{Step 1}: 
Note that $(a_1,a_n)$ is the unique $\widehat{\succsim}_l$-greatest 
element; hence top-ranked in $\wintenl$ too.
\vspace{3pt}

\noindent
\textit{Step 2}: By \eqref{PIF3}, \eqref{elicit-eq1} and $\succ_l$, 
any pair $(a_i,a_j)$ other than $(a_1,a_{n-1})$ and $(a_2,a_n)$
is ranked below one of these two pairs under $\widehat{\succsim}$. 
Moreover, $(a_1,a_{n-1})$ and $(a_2,a_n)$ are incomparable 
by $\widehat{\succsim}_l$. 
Assign the second place in $\widehat{\succsim}_l$ 
to one of these pairs after a binary query.
\vspace{3pt}

\noindent
\textit{Step 3}: If $(a_1,a_{n-1})$ 
is placed second in Step 2, then, by the 
same logic, assign the third place to 
$(a_2,a_n)$ or $(a_1,a_{n-2})$ after a binary query.
If, instead, $(a_2,a_n)$ is placed second in Step 2, 
assign the third place to $(a_1,a_{n-1})$ or $(a_3,a_n)$
after a binary query.
\vspace{3pt}

$\vdots$
\vspace{3pt}

\noindent
\textit{Step} $k$: 
Select one pair from the remaining $\widehat{\succsim}_l$-maximal 
ones and assign it the $k$th place.
\vspace{3pt}

$\vdots$
\vspace{3pt}

\noindent
\textit{Step} $\binom{n}{2}-1$: Observe that, 
by \eqref{PIF3}, each of the last two 
$\widehat{\succsim}_l$-unranked pairs is necessarily of the form 
$(a_i,a_{i+1})$ and $(a_j,a_{j+1})$.
The last (binary) query selects one for 
the $\left(\binom{n}{2}-1\right)$th position in $\wintenl$.
\vspace{3pt}

\noindent
\textit{Step} $\binom{n}{2}$: The last remaining pair is assigned 
the bottom place in $\wintenl$.
\vspace{3pt}

\noindent
By construction, finally, $\wintenl$ induces $\succ_l$.
\hfill $\blacksquare$\\

\begin{figure}[!htbp]
\centering
\caption{Decision tree of the ordinal intensities elicitation algorithm}
\begin{tikzpicture}[font=\footnotesize,edge from parent/.style={draw,thick}]
% Two node styles: solid and hollow
\tikzstyle{solid node}=[circle,draw,inner sep=1.2,fill=black];
\tikzstyle{hollow node}=[circle,draw,inner sep=1.2];

% The initial node
\node(start)[hollow node]{};

% The left branch
\node(s1a)[below left = -0.8cm and -0.8cm of start]{\small$(a_1,a_n)$};
\node(a)[below left = 2cm and 2.3cm of start]{\small$(a_1,a_{n-1})$};

\node(aa1)[below left = 2.5cm and -0.5cm of a]{\small$(a_1,a_{n-2})$};
\node(aa3)[below left = 1.92cm and -0.2cm of aa1]{\small$(a_1,a_{n-3})$};
\node(aa4)[below right = 1.92cm and -1.8cm of aa1]{\small$(a_2,a_n)$};
\node(aa3lab)[below left = 0.5cm and -0.3cm of aa1]{};
\node(PAa1b2)[below right = 2.5cm and -0.5cm of a]{$(a_2,a_n)$};

\node(q1a1)[below left = 0.7cm and -0.1cm of PAa1b2]{};
\node(q'1a1)[below right = 0.65cm and -0.25cm of PAa1b2]{};
\node(ab)[below left = 2cm and -0.7cm of PAa1b2]{\small$(a_1,a_{n-2})$};
\node(bb2)[below right = 2cm and -0.7cm of PAa1b2]{\small $(a_2,a_{n-1})$};

\node(b)[below right = 2cm and 2cm of start]{$(a_2,a_n)$};
\node(PAb1a2)[below left = 2.5cm and -0.8cm of b]{$(a_1,a_{n-1})$};
\node(bb1)[below right = 2.5cm and 0.9cm of b]{$(a_3,a_n)$};
\node(bb3)[below right = 2cm and -0.5cm of bb1]{\small$(a_1,a_{n-1})$};
\node(bb4)[below left = 2cm and -0.8cm of bb1]{\small$(a_4,a_n)$};
\node(bb3lab)[below right = 0.65cm and -0.2cm of bb1]{\small};	
\node(ba)[below left = 2cm and -0.8cm of PAb1a2]{\small$(a_1,a_{n-2})$};
\node(aa2)[below right = 2cm and -0.7cm of PAb1a2]{\small$(a_3,a_n)$};

\node(prefinalleft)[below left = 2.5cm and -0.7cm of bb2]{\small$(a_i,a_{i+1})$};
\node(prefinalmiddle)[below right = 2.5cm and -0.3cm of bb2]{\small$(a_k,a_{k+1})$};
\node(prefinalright)[below right = 2.5cm and 2.7cm of bb2]{\small$(a_j,a_{j+1})$};

\node(finalleft)[below left = 2cm and -0.7cm of prefinalmiddle]{\small$(a_i,a_{i+1})$};
\node(finalright)[below right = 2cm and 0cm of prefinalmiddle]{\small$(a_j,a_{j+1})$};

\node(last)[below right = 1cm and -1cm of finalleft]{\small$(a_j,a_{j+1})$};

% The period label nodes
\node(t=1)[below right = 0cm and 7.5cm]{\small
\textbf{Step 1}};
\node(t=2)[below right = 2cm and 7.5cm]{\small
\textbf{Step 2}};
\node(t=3)[below right = 5.1cm and 7.5cm]{\small
\textbf{Step 3}};
\node(t=4)[below right = 7.8cm and 7.5cm]{\small
\textbf{Step 4}};
\node(t=5)[below right = 11cm and 6.5cm]{\small
\textbf{Step $\binom{n}{2}-2$}};	
\node(t=6)[below right = 13.5cm and 6.5cm]{\small
\textbf{Step $\binom{n}{2}-1$}};
\node(t=7)[below right = 15.5cm and 7.5cm]{\small
	\textbf{Step $\binom{n}{2}$}};

% Some vertical dots
\node(vdots1)[below right = 8.5cm and -0.2cm]{\small$\vdots$};
\node(vdots2)[below right = 9cm and -0.2cm]{\small$\vdots$};

% Specify links
\draw[very thick, ->] (start) edge (a);
\draw[very thick, ->] (start) edge (b); 
\draw[very thick, ->] (a) edge (aa1);
\draw[very thick, ->] (aa1) edge (aa3);
\draw[very thick, ->] (aa1) edge (aa4);
\draw[very thick, ->] (a) edge (PAa1b2);
\draw[very thick, ->] (b) edge (PAb1a2);
\draw[very thick, ->] (b) edge (bb1);
\draw[very thick, ->] (bb1) edge (bb3);	
\draw[very thick, ->] (bb1) edge (bb4);	
\draw[very thick, ->] (PAa1b2) edge (ab);
\draw[very thick, ->] (PAa1b2) edge (bb2);
\draw[very thick, ->] (PAb1a2) edge (ba);
\draw[very thick, ->] (PAb1a2) edge (aa2);
\draw[very thick, ->] (vdots2) edge (prefinalleft);
\draw[very thick, ->] (vdots2) edge (prefinalmiddle);
\draw[very thick, ->] (vdots2) edge (prefinalright);
\draw[very thick, ->] (prefinalmiddle) edge (finalleft);
\draw[very thick, ->] (prefinalmiddle) edge (finalright);
\draw[very thick, ->] (finalleft) edge (last);
\end{tikzpicture}
\label{fig:elicit}
\end{figure}

\begin{example}
We illustrate this algorithm and explain its 
accompanying decision tree in Figure \ref{fig:elicit} 
with the intensity ordering 
and preferences of the first agent in Example 2: 
$(a_1,a_4)\sinten_1(a_2,a_4)\sinten_1(a_1,a_3)\sinten_1(a_2,a_3)
\sinten_1(a_1,a_2)\sinten_1(a_3,a_4)$ and
$a_1\succ_1 a_2 \succ_1 a_3 \succ_1 a_4$. In line with 
Proposition \ref{prp:elicit}, we assume that $\succ_1$ has already
been elicited. The construction of $\sinten_1$ from $\succ_1$ 
proceeds as follows:

\noindent
\textit{Step 1:} Assign $(a_1,a_4)$ in the first place.

\noindent
\textit{Step 2/Query 1:} 
Assign $(a_1,a_3)$ or $(a_2,a_4)$ \textnormal{(choice)}
in the second place. 

\noindent
\textit{Step 3/Query 2:} 
Assign $(a_1,a_3)$ \textnormal{(choice)} 
or $(a_3,a_4)$ in the third place. 

\noindent
\textit{Step 4/Query 3:} 
Assign $(a_1,a_2)$, $(a_2,a_3)$ \textnormal{(choice)}
or $(a_3,a_4)$ in the fourth place.

\noindent
\textit{Step 5/Query 4:} 
Assign $(a_1,a_2)$ \textnormal{(choice)} or 
$(a_3,a_4)$ in the fifth place.

\noindent
\textit{Step 6:} 
Assign $(a_3,a_4)$ in the sixth place.	
\hfill $\blacklozenge$
\end{example}

We now turn to the problem of using a profile of 
elicited strict intensity rankings, together with the 
strict preference orderings nested within them, 
to find an allocation that improves 
upon the Pareto-efficiency criterion.
We do so by building on, and extending, the classic Random Priority 
mechanism (RP; also known as \textit{random serial dictatorship}). 
This involves ranking agents according to a uniform-randomly 
drawn priority order and inviting each agent, in that order,  
to select their most preferred item among those not yet 
chosen by higher-ranked agents. 
RP is strategy-proof and leads to a(n) (ex-post)
Pareto-efficient allocation (see 
\citeauthor{abdulkadiroglu&sonmez98}, 
\citeyear{abdulkadiroglu&sonmez98}).
\cite{pycia-troyan26} recently showed that RP 
is, in fact, the \textit{unique} mechanism with this Pareto-efficiency 
property that is also \textit{symmetric} (ex-ante fair) as well as 
\textit{obviously} strategy-proof in the sense of \cite{li17}.

We are interested in a matching rule that guarantees 
Pareto-efficiency of the final allocation and also ensures 
that any potential welfare improvements that could be materialized 
by application of the intensity-dominance 
criterion do, in fact, materialize. 
It is because of its simplicity and desirable 
efficiency, fairness and incentive properties that we take RP 
as our starting point here, and extend it so as to incorporate 
agents' intensity rankings. More specifically, we consider 
the following matching algorithm:

\vspace{8pt}

\noindent \textit{Input:} 
A strict intensity profile $S$, canonically represented by $s$.

\vspace{3pt}

\noindent \textit{Step 1:} 
Choose a linear order $\gg_1$ over $N$ from the 
uniform distribution with support $n!$.

\vspace{3pt}

\noindent \textit{Step 2:} 
Apply RP under $\gg_1$ on the induced preference profile $P_S$ 
to arrive at a Pareto-efficient allocation $x$.

\vspace{3pt}

\noindent \textit{Step 3:}
Choose a linear order $\gg_2$ over $N$ from the 
uniform distribution with support $n!$.

\vspace{3pt}

\noindent \textit{Step 4:}
Denoting by $j_{\gg_2}$ the agent who is ranked 
$j$th under $\gg_2$, search for the $\gg_2$-lowest agent 
$j\in N\setminus\{1_{\gg_2}\}$ such that 
$s_1{_{_{\gg_2}}}(x_j{_{_{\gg_2}}},x_1{_{_{\gg_2}}})
>s_j{_{_{\gg_2}}}(x_j{_{_{\gg_2}}},x_1{_{_{\gg_2}}})$. 
If such an agent exists, swap $x_1{_{_{\gg_2}}}$ 
with $x_j{_{_{\gg_2}}}$
and remove agents $1{_{_{\gg_2}}}$, 
$j{_{_{\gg_2}}}$ from $N$ and items 
$x_1{_{_{\gg_2}}}$, $x_j{_{_{\gg_2}}}$ from $A$.\\
If no such agent exists, 
remove $1{_{_{\gg_2}}}$ and $x_1{_{_{\gg_2}}}$.

\vspace{3pt}

\noindent \textit{Repeat} Step 4 until no agents remain.  

\vspace{8pt}

The first two steps set up the RP algorithm.
The third step draws a new priority ordering over agents,
which is generally different from the one drawn earlier in the RP stage. 
The fourth step starts with the agent ranked first under 
this new priority order and searches---without loss, 
from the bottom up---for an agent whose RP assignment 
is preferred by the first-ranked agent to their own assignment 
\textit{more} than it is so by lower-ranked agent.
If such an individual exists, the algorithm implements the swap, 
removes the two agents and the respective items, then 
repeats this step with the remaining agents and items.
If not, then top-ranked agent and their RP endowment 
are removed before the step is re-applied. 
This forced exit is important in two ways.
First, it prevents potentially indefinite loops; for example
in cases where the RP allocation is already intensity-efficient. 
Second, as is clarified below, it ensures that 
the computational complexity of the whole process
falls in the manageable quadratic class.

\begin{prp}
\label{prp:complexity}
The algorithm ends in at most ${{n}\choose{2}}$ 
steps---hence in $O(n^2)$ time---and produces 
the Pareto-efficient allocation of its RP-implementing 
step or an allocation that intensity-dominates it.
\end{prp}

In general, this algorithm does not produce an intensity-efficient 
allocation. This is so because the forced exit of agents and items 
in the fourth step means that not all pairwise possibilities 
for intensity-dominance driven improvements are considered in 
the transition from the RP-mandated Pareto-efficient allocation 
in Step 2 to the final allocation. 
Clearly, however, the algorithm improves upon RP and 
Pareto-efficiency---even if potentially partially so---in the direction 
of intensity-efficiency, without adding intractable computational 
complexity. 
Importantly, moreover, what is perhaps not immediately obvious 
is that welfare improvements in the intensity-dominance direction are
possible even if \textit{no} intensity-efficient allocation exists
at the input profile.

\begin{cor}
There are profiles with no intensity-efficient allocation at which 
the algorithm can lead to a final allocation that intensity-dominates
the Pareto-efficient one of its interim RP stage.
\end{cor}

To see this, consider the profile described in 
Table \ref{tab:counterexample4}, where, 
as was explained as part of the discussion 
of Theorem \ref{thm:n=3}, no intensity-efficient 
allocation exists. 
All four agents prefer $a$ to $b$ to $c$ to $d$.
Suppose that the first randomly drawn priority ordering is 
$1\gg_1 2 \gg_1 4 \gg_1 3$. In Step 2 (RP) of the algorithm  
this leads to the Pareto-efficient allocation $x:=(a,b,d,c)$.
Suppose that the second priority ordering, drawn randomly 
in Step 3, is $1\gg_2 2 \gg_2 3 \gg_2 4$. 
In Step 4, the agent-item pairs $(1,a)$ (first) 
and $(2,b)$ (second) exit with their respective 
RP assignments in $x$. 
In the last iteration of that step, however, agent 3---who
is ranked third in $\gg_2$---swaps items with agent 4, because
$s_3(c,d)=2>1=s_4(c,d)$.

It is presently unclear how this algorithm can be modified 
to deliver an intensity-efficient 
allocation---when one exists---without trivially 
requiring factorial search 
over all possible priority orderings over agents in Steps 1 and/or 3.
That said, the potentially even stronger welfare gains that 
can be associated with such more exhaustive exploration 
do not come with considerable computational burdens in 
``small-market'' one-to-one assignments such as dormitory rooms to 
students; classrooms to courses; offices or 
work shifts to workers.\footnote{For cases where $n\leq 6$ 
we can confirm that brute-force search algorithms instantaneously 
find and list all intensity-efficient allocations. Details are 
available from the author upon request.}

One is naturally interested to know if, 
in addition to being incentivized to report their preferences 
truthfully by virtue of the RP part of the algorithm, agents
are also incentivized to report their intensities truthfully
in this environment. 
The answer is negative. 
To see this, suppose $n=3$ and let $S$
be such that $a\succ_1 b\succ_1 c$; $a\succ_2 b\succ_2 c$; 
$c\succ_3 b \succ_3 a$; $(a,b)\sinten_1 (b,c)$; $(b,c)\sinten_2 (a,b)$;  
$(c,b)\sinten_3(b,a)$. The possible orderings over $N$ that 
might be drawn in Steps 1 and 3 are: 
(i) $1\gg 2 \gg 3$; (ii) $1\gg 3 \gg 2$; (iii) $2\gg 1 \gg 3$;
(iv) $2\gg 3 \gg 1$; (v) $3\gg 1 \gg 2$; (vi) $3\gg 2 \gg 1$.
The two Pareto-efficient allocations here are $(a,b,c)$, $(b,a,c)$,  
and the former intensity-dominates the latter. 
Suppose the algorithm's second step delivers this 
intensity-efficient allocation
(under RP, this happens with prob. $\frac{1}{2}$).
If intensities are reported truthfully, there is no order $\gg$
under which the algorithm's subsequent steps alter it.  
But if $(a,b)\sinten'_2 (b,c)$ is the second agent's misreported 
intensity, then under the $\gg_2$-orders (iii), (iv) and (vi)---hence 
with probability $\frac{1}{2}$---the algorithm reassigns $b$ to $1$
and $a$ to $2$, making the latter agent better off.

Interestingly, the incentives to misreport one's 
\textit{intensities} in this way, namely toward increasing one's 
chances of getting their most preferred alternative, act in the 
opposite direction to the incentives toward misreporting one's 
\textit{preferences} in the Boston/Immediate-Acceptance 
mechanism of the distinct two-sided (School Choice) matching problem. 
Specifically, as has been discussed by 
\citet*{pathak-sonmez08,abdulkadirogluetal11,kojima-unver14}
and other authors, under this mechanism students may (and often do)
submit their second- or third-ranked school as their supposedly 
most preferred one if their actual top-ranked school is very popular, 
out of concern that in the event they are not assigned to it they 
may end up at one of their lower-ranked schools if their second 
or third most preferred ones have in the meantime filled up.
\cite{abdulkadirogluetal11}, however, as well as
\cite{miralles09}, \cite{pycia11}, 
\cite{featherstone-niederle16} and other studies, 
argued that this drawback on 
incentive grounds of the Boston mechanism may be outweighed 
by the total welfare gains associated with such misreporting once
the students' intensities are accounted for.
In line with the prior relevant literature, 
intensity comparisons in that paper's Bayesian-Nash equilibrium 
analysis are derived from intra- and inter-personal differences 
in cardinal utility functions that emerge from expected-utility preferences
over lotteries over schools.
In that framework too, finally, the intensity information 
that is conveyed to the social planner is assumed to exist 
without having been elicited incentive-compatibly, 
and it is unclear if such elicitation is possible. 
In relation to this, \cite{pycia-unver25} recently showed that, 
in environments without monetary transfers such as the 
one considered above, only ordinal 
mechanisms are \textit{group} strategy-proof.

\section{Concluding Remarks}

In this paper we study the classic one-to-one assignment problem 
in an environment where, in addition to standard ordinal preferences, 
agents are also assumed to have ordinal preference intensities. 
These are internally consistent comparisons between preference 
improvements that do not rely on any quantification
of the kind implied by cardinal (or pseudo-cardinal) utility functions.
Derived as it is from binary comparisons rather than score 
assignments, such intensity information is less cognitively demanding 
and, for that reason, potentially more reliable.

An interpersonal comparability assumption is introduced 
for such ordinal intensities, contrasting across agents the relative 
ranking of an improvement from one alternative to another. 
In the very special case where agents' intensity orderings 
are defined by value differences with respect to a cardinal utility 
function, this assumption is distinct from standard interpersonal 
comparability of utility differences that underpins 
cardinal-welfarist criteria such as classical or 
relative utilitarianism. 

Building on this assumption, 
we introduce and analyse the concept of intensity-efficient allocations. 
These are based on an intuitive dominance criterion 
and refine the set of Pareto-efficient allocations in ways 
that incorporate into the allocation process such ordinal 
intensity information with the goal---shared by notions defined 
in the cardinal-welfarist tradition---of achieving 
greater welfare gains and distributive justice. 
We identify necessary and sufficient 
conditions for the general existence of such allocations.
While distinct, these conditions revolve around the concept of 
homeo-monotonic ordinal intensity profiles, whose central idea
is that, whenever agents whose preferences are identical up to a point 
can be ordered in terms of the intensity with which they prefer 
any two consecutively-ranked items, then agents can be ordered 
in the same way in terms of how intensely they prefer items 
that are not ranked consecutively.  

We show, finally, how this paper's novel analytical environment 
for social choice could be applied in matching problems. 
In particular, we devise an algorithm that takes as input  
an agent's preference ordering and, utilizing the structural 
properties of the intensity orderings in the class under study, 
elicits such an ordering in strictly fewer steps (in fact, 
quadratically fewer) than what naive binary querying would achieve. 
We then offer an extension of the classic---and with several desirable
properties---Random Priority mechanism that uses ordinal intensity 
profiles as its input and returns an allocation that either coincides 
with the Pareto-efficient one that is achieved under Random Priority
or one that intensity-dominates it. Notably, such welfare 
improvements over the benchmark mechanism are possible 
under its proposed extension even in rare cases where 
the input intensity profile does not have an intensity-efficient
allocation.

To sum up, this paper sets up in detail a new conceptual,   
methodological and analytical environment for the study 
of allocation problems, providing a middle ground between 
purely ordinal and fully cardinal ones. 
We hope that this ground will prove fertile in theoretical 
and practical applications. 

\appendix

\section*{Appendix: Remaining Proofs}

\noindent \textbf{\textit{Proof of Lemma \ref{lem:sincov}.}}
Upon viewing $s_l$ as a general function 
of two variables, \eqref{PIF4} is known as \textit{Sincov's} 
functional equation, whose solution is some $f:A\rightarrow\mathbb{R}$,
unique up to an additive constant, such that 
$s_l(a,b)\equiv f(a)-f(b)$ \citep{aczel}. 
This establishes the ``if'' part. 
The ``only if'' part is immediate upon defining 
$s_l(a,b):=u_l(a)-u_l(b)$ for the postulated 
function $u_l$ with the requisite representation property.
\hfill $\square$\\

\noindent \textbf{\textit{Proof of Lemma \ref{lem:canonical}.}}
Let $s_l$ be an arbitrary preference 
intensity function for $\wintenl\in\mathcal{I}$. 
Define the level set of $s_l$ at $(a,b)\in A\times A$ 
by $\mathbf{s}_l[a,b]:=\{(a',b')\in A\times A: s_l(a,b)=s_l(a',b')\}$.
Further, define $r_l:A\times A\rightarrow \R$ by 
\begin{eqnarray*}
r_l(a,b) & := & 
\left\{
\begin{array}{ll}
|\{\mathbf{s}_l[a',b']: s_l(a,b)>s_l(a',b')\geq s_l(e,e)\}|, 
& \text{if } s_l(a,b)>s_l(e,e)\\
0, & \text{if } s_l(a,b)=s_l(e,e)\\
-r_l(b,a), & \text{if } s_l(a,b)<s_l(e,e)
\end{array}
\right.
\end{eqnarray*}	
By construction, $r_l$ is canonical and 
$r_l(a,b)\geq r_l(c,d)\Leftrightarrow 
s_l(a,b)\geq s_l(c,d)$.	
\hfill $\square$\\

\noindent \textbf{\textit{Proof of Theorem \ref{thm:n=3}.}}
The argument proceeds by considering the possible ways 
in which an arbitrary profile $S\in\mathcal{S}$ when $n=3$
might generate a sequence of distinct Pareto-efficient allocations 
that are implicated in an intensity-dominance cycle. 
To this end, let $D$ be the intensity-dominance relation 
that is introduced in Definition \ref{intensity-dominance}. 
Suppose to the contrary that 

\vspace{-15pt}

\begin{equation}
\label{sequence} w^1Dw^2D\ldots Dw^kDw^1
\end{equation}
for Pareto efficient allocations $w^1,\ldots,w^k$ on $X:=\{a,b,c\}$. 

\begin{obs}
\label{FirstThmObs1} 
$n=3$ implies that for any two allocations $w^i,w^{i+1}$ 
such that $w^iDw^{i+1}$ it must be that $w^i_l=w^{i+1}_l$ 
for exactly one agent $l\in\{1,2,3\}$ and 
$(w^i_j,w^i_k)=(w^{i+1}_k,w^{i+1}_j)$ for $j,k\neq l$.	
\end{obs}

\begin{obs}
\label{FirstThmObs2} $n=3$ implies $k\leq 6$. 	
\end{obs}

\begin{obs}
\label{FirstThmObs3} Pareto efficiency of $w^i=(a',b',c')$ and Strictness 
together imply 
\begin{eqnarray}
\label{PE1}	s_2(a',b')>0 & \Longrightarrow & s_1(a',b')>0\\
\label{PE2} s_3(b',c')>0 & \Longrightarrow & s_2(b',c')>0\\
\label{PE3}	s_3(a',c')>0 & \Longrightarrow & s_1(a',c')>0
\end{eqnarray}
\textnormal{Indeed, $w^i$ could be Pareto-improved upon 
if any of these implications was false.}
\end{obs}

\begin{obs}
\label{FirstThmObs4} Strictness and canonicality 
of $s_i$ 
imply $s_i(a',b')=s_i(c',d')>0 \Leftrightarrow (a',b')=(c',d')$ 
and, jointly with $n=3$, also 
$s_i(a',b')>0 \Leftrightarrow s_i(a',b')\in \{1,2,3\}$.
\end{obs}

Notice that \eqref{sequence} is impossible for $k=2$ because 
$D$ is asymmetric by construction. Suppose $k=3$. 
Without loss of generality, write $w^1:=(a,b,c)$ and $w^2:=(b,a,c)$. 
Then, by Observation \ref{FirstThmObs1}, 
either $w^3 = (b,c,a)$ or $w^3=(c,a,b)$. 
Since, in both cases, $w^1$ and $w^3$ are $D$-incomparable by construction, 
the $w^3Dw^1$ postulate in \eqref{sequence} is contradicted.

Now suppose $k=4$. By \eqref{sequence} and the above implications, 
we may take $w^1$, $w^2$, $w^3$ 
to be as in the $k=3$ case, from which it then follows 
that allocation $w^4$ must be either 
$(c,b,a)$ or $(a,c,b)$. 
Notice that either possibility 
is compatible with $w^3=(b,c,a)$ 
and with $w^3=(c,a,b)$. 
We therefore have the following 4 cases to consider:

\noindent $[$\textit{Note:} in what follows 
we make repeated use---often without explicit reference---of 
the Pareto-efficiency implications \eqref{PE1}--\eqref{PE3}, 
the skew-symmetry property $s_l(a,b)=-s_l(b,a)$ of canonical 
representations and, whenever exact values of the $s_i$ functions are asserted, 
of the lateral-consistency property \eqref{PIF3} 
together with the assumption that every $s_i$ is canonical and strict 
(cf Observations \ref{FirstThmObs3}--\ref{FirstThmObs4}).$]$

\noindent\textit{Case 1.} $w^3=(b,c,a)$, $w^4=(c,b,a)$. 
By the definition of $D$, and by the above assumptions:
\vspace{-10pt}
\begin{eqnarray}
\label{PE-1} w^1Dw^2 & \Longrightarrow & s_1(a,b)>s_2(a,b)\\
\label{PE-2} w^2Dw^3 & \Longrightarrow & s_2(a,c)>s_3(a,c)\\
\nonumber w^3Dw^4 & \Longrightarrow & s_1(b,c)>s_2(b,c)\\
\nonumber w^4Dw^1 & \Longrightarrow & s_3(a,c)>s_1(a,c)
\end{eqnarray}
Therefore,
\begin{equation}
\label{PEcase1}	s_2(a,c)>s_3(a,c)>s_1(a,c)
\end{equation} 
Since $w^i$ is Pareto efficient for $i\leq 4$, it follows 
from \eqref{PE1}--\eqref{PE3} that there are 4 subcases to consider:\\
\textit{Subcase 1-i}. $s_1(a,b)>s_2(a,b)>0$ and $s_1(b,c)>s_2(b,c)>0$. 
By \eqref{PIF3} and the fact that $s_1,s_2$ are canonical, 
this implies $s_1(a,c)=3$, which contradicts \eqref{PEcase1}.\\
\textit{Subcase 1-ii}. $s_1(a,b)>s_2(a,b)>0$ and $s_2(c,b)>s_1(c,b)>0$. 
Suppose $s_1(a,c)>0$ is also true. Then, by \eqref{PIF3}, \eqref{PEcase1} 
and the fact that $s_1$ is canonical, $s_1(a,c)=1$ and $s_1(a,b)=3$. 
If $s_2(a,c)>0$ is also true, then $s_2(a,b)=3$. 
This contradicts $s_1(a,b)>s_2(a,b)$. 
So, it must be that $s_2(c,a)>0$ instead. 
But in this case $s_2(c,a)>0$, $s_2(a,b)>0$ and \eqref{PIF3} together 
imply $s_2(c,b)=3$. 
This contradicts $s_2(a,c)=3$ which is now implied by 
\eqref{PEcase1} and the fact that the profile $s$ is canonical. 
Thus, it must be that $s_1(c,a)>0$ instead. 
So now we have $s_1(c,a)>0$, $s_1(a,b)>0$, which implies $s_1(c,b)=3$. 
But since, by assumption, $s_2(c,b)>s_1(c,b)$ and $s_2$ is canonical, 
this is a contradiction.\\ 
\textit{Subcase 1-iii}. $s_2(b,a)>s_1(b,a)>0$ and $s_1(b,c)>s_2(b,c)>0$. 
Suppose first that $s_2(a,c)>0$ is also true. 
Then, $s_2(b,a)>0$ and $s_2(a,c)>0$ implies $s_2(b,c)=3$. 
If $s_1(a,c)>0$ is also true, then \eqref{PEcase1} and the fact that 
$s$ is canonical together imply $s_2(a,c)=s_2(b,c)$, which contradicts 
\eqref{PIF3} and Strictness. 
So, it must be that $s_1(c,a)>0$. 
From $s_1(b,c)>0$ and $s_1(c,a)>0$ we now get $s_1(b,a)=3$. 
In view of $s$ being canonical, this contradicts $s_2(b,a)>s_1(b,a)$.\\ 
\textit{Subcase 1-iv}. $s_2(b,a)>s_1(b,a)>0$ and 
$s_2(c,b)>s_1(c,b)>0$. 
Because $s$ is canonical, this and \eqref{PIF3} readily imply 
$s_1(c,a)=s_2(c,a)=3$. 
But since \eqref{PEcase1} is equivalent to $s_1(c,a)>s_3(c,a)>s_2(c,a)$, 
this is a contradiction.

\noindent Hence, $w^4Dw^1$ is impossible for such $w^3$ and $w^4$.

\noindent	\textit{Case 2.} $w^3=(c,a,b)$, $w^4=(c,b,a)$.\\ 
We again have $w^1Dw^2 \Longrightarrow s_1(a,b)>s_2(a,b)$ 
and $w^2Dw^3 \Longrightarrow s_1(b,c)>s_3(b,c)$ and, in addition, 
\begin{eqnarray}
\label{PEcase2c}	w^3Dw^4 & \Longrightarrow & s_2(a,b)>s_3(a,b)\\
\label{PEcase2d}	w^4Dw^1 & \Longrightarrow & s_3(a,c)>s_1(a,c)
\end{eqnarray}
Therefore,
\begin{equation}
\label{PEcase2} s_1(a,b)>s_2(a,b)>s_3(a,b)
\end{equation}
In view of \eqref{PE1}--\eqref{PE3}, we can now consider 
the following 4 possible subcases:\\
\textit{Subcase 2-i.} $s_1(a,b)>s_2(a,b)>0$ and $s_1(b,c)>s_3(b,c)>0$. 
By \eqref{PEcase2} and \eqref{PE1}--\eqref{PE3}, 
the former postulate implies $s_3(a,b)>0$. 
Since $s$ is canonical, this further implies $s_3(a,b)=1$, $s_2(a,b)=2$ 
and $s_1(a,b)=3$. 
This, canonicality of $s$ and $s_1(b,c)>s_3(b,c)>0$ together imply 
$s_1(b,c)=2$ and $s_3(b,c)=1=s_3(a,b)$, which contradicts Strictness.\\
\textit{Subcase 2-ii.} $s_1(a,b)>s_2(a,b)>0$ and $s_3(c,b)>s_1(c,b)>0$. 
For the same reasons as in 2-i, we have $s_3(a,b)=1$, $s_2(a,b)=2$ 
and $s_1(a,b)=3$. 
This, together with canonicality of $s$ and $s_3(c,b)>s_1(c,b)>0$, 
further implies $s_1(c,b)=1$. 
Hence, it also follows that either $s_1(a,c)=2$ or $s_1(c,a)=2$. 
The latter possibility cannot be valid, for \eqref{PIF3} and $s_1(c,a)>0$, 
$s_1(a,b)>0$ would then imply $s_1(c,b)=3$, which contradicts $s_1(c,b)=1$. 
Consider then the case of $s_1(a,c)=2$. 
This, together with \eqref{PEcase2d} and canonicality of $s$, 
implies $s_3(a,c)=3$. 
Thus, we have $s_3(a,c)=3$, $s_3(a,b)=1$ and, from $s_3(c,b)>s_1(c,b)>0$ 
and canonicality, $s_3(c,b)=2$. 
But, by \eqref{PIF3} and canonicality, $s_3(a,c)>0$ and $s_3(c,b)>0$ 
implies $s_3(a,b)=3$, a contradiction.\\
\textit{Subcase 2-iii.} $s_2(b,a)>s_1(b,a)>0$ and $s_1(b,c)>s_3(b,c)>0$. 
The former postulate, together with \eqref{PEcase2} and canonicality, 
implies $s_3(b,a)=3$, $s_2(b,a)=2$, $s_1(b,a)=1$. 
By \eqref{PEcase2c}, either $s_3(a,c)>s_1(a,c)>0$ or 
$s_1(c,a)>s_3(c,a)>0$ also holds. Consider the first possibility. 
From $s_1(b,a)=1$, $s_1(a,c)>0$, \eqref{PIF3} and canonicality we get 
$s_1(a,c)=2$. This and \eqref{PEcase2d} implies $s_3(a,c)=3$. 
Since $s_3(b,a)=3$ is also true, this contradicts Strictness. 
Hence, it must be that $s_1(c,a)>s_3(c,a)>0$. 
But in this case $s_1(b,c)>0$, $s_1(c,a)>0$, \eqref{PIF3} 
and canonicality imply $s_1(b,a)=3$, which contradicts \eqref{PEcase2}.\\
\textit{Subcase 2-iv.} $s_2(b,a)>s_1(b,a)>0$ and $s_3(c,b)>s_1(c,b)>0$. 
As in 2-iii, we have $s_3(b,a)=3$, $s_2(b,a)=2$, $s_1(b,a)=1$. 
But $s_3(c,b)>0$, $s_3(b,a)>0$ and \eqref{PIF3} imply $s_3(c,a)>s_3(b,a)=3$ which, 
by canonicality, is impossible.

\noindent Hence, $w^4Dw^1$ is impossible for such $w^3$ and $w^4$ too.

\noindent\textit{Case 3.} $w^3=(b,c,a)$, $w^4=(a,c,b)$.\\ 
We now have
\begin{eqnarray}
w^3Dw^4 & \Longrightarrow & s_3(a,b)>s_1(a,b)\\
\label{PEcase3d}	w^4Dw^1 & \Longrightarrow & s_3(b,c)>s_2(b,c)
\end{eqnarray}
It follows that
\begin{equation}
\label{PEcase3} s_3(a,b)>s_1(a,b)>s_2(a,b)
\end{equation}
We consider the 4 subcases that are now possible:\\
\textit{Subcase 3-i.} $s_1(a,b)>s_2(a,b)>0$ and $s_1(b,c)>s_3(b,c)>0$. 
The latter, together with \eqref{PE1}--\eqref{PE3}, \eqref{PEcase3d} 
and canonicality, 
implies $s_3(a,b)=1$. But canonicality and \eqref{PE1}--\eqref{PE3} 
also implies $s_3(a,b)=3$, which contradicts Strictness.\\
\textit{Subcase 3-ii.} $s_1(a,b)>s_2(a,b)>0$ and $s_3(c,b)>s_1(c,b)>0$. 
The first postulate and \eqref{PEcase3}, together with canonicality, 
implies $s_3(a,b)=3$, $s_1(a,b)=2$ and $s_2(a,b)=1$. 
Since $s_3(c,b)>s_1(c,b)>0$ is also assumed, this and canonicality further 
imply $s_3(c,b)=2$. Now, because $s_3(a,b)>0$ and $s_3(c,b)>0$, 
it follows from \eqref{PIF3} that $s_3(c,a)>0$ too. 
But \eqref{PIF3} in this case further implies $s_3(c,a)>s_3(a,b)=3$, 
which is impossible.\\
\textit{Subcase 3-iii.} $s_2(b,a)>s_1(b,a)>0$ and $s_1(b,c)>s_3(b,c)>0$. 
The first postulate, together with \eqref{PEcase3} and canonicality, 
implies $s_2(b,a)=3$, $s_1(b,a)=2$, $s_3(b,a)=1$. 
The second postulate and $s_1(b,a)=2$, 
together with Strictness, implies $s_1(b,c)=3$. 
This in turn implies $s_1(a,c)=1$ or $s_1(c,a)=1$. 
If the latter is true, then $s_1(b,c)>0$, $s_1(c,a)>0$ and \eqref{PIF3}, 
together with canonicality, implies $s_1(b,a)=3$, a contradiction. 
Hence, it must be that $s_1(a,c)=1$. 
We therefore have $s_1(b,a)=2$, $s_1(a,c)=1$ and, by \eqref{PIF3} 
and canonicality, $s_1(b,c)=3$. 
From \eqref{PE1}--\eqref{PE3}, \eqref{PE-2}, \eqref{PEcase3d}
and canonicality we also know that $s_1(b,c)>s_3(b,c)>s_2(b,c)>0$ implies 
$s_3(b,c)=2$ and $s_2(b,c)=1$. 
Thus, we have $s_3(b,a)>0$, $s_3(b,c)>0$ and, by \eqref{PE1}--\eqref{PE3} and 
$s_1(a,c)>0$, 
also $s_3(a,c)>0$. But $s_3(b,a)>0$, $s_3(a,c)>0$ together with \eqref{PIF3} 
and canonicality implies $s_3(b,c)=3$, a contradiction.\\
\textit{Subcase 3-iv.} $s_2(b,a)>s_1(b,a)>0$ and $s_3(c,b)>s_1(c,b)>0$. 
These readily imply $s_1(c,a)=3$. 
As above, \eqref{PEcase3} implies $s_2(b,a)=3$, $s_1(b,a)=2$ and $s_3(b,a)=1$. 
By \eqref{PIF3}, $s_1(c,a)=3$ and $s_1(a,b)=2$ implies $s_1(c,b)=1$. 
From the above postulates and from \eqref{PEcase3d}, $s_2(c,b)>s_3(c,b)>s_1(c,b)$ 
further implies $s_2(c,b)=3$, which contradicts $s_2(b,a)=3$ and Strictness.

\noindent Hence, $w^4Dw^1$ is impossible for such $w^3$ and $w^4$ here as well.

\noindent	\textit{Case 4.} $w^3=(c,a,b)$, $w^4=(a,c,b)$.  
It is now true that 
\begin{eqnarray}
\label{PEcase4c} w^3Dw^4 & \Longrightarrow s_2(a,c)>s_1(a,c)\\
w^4Dw^1 & \Longrightarrow s_3(b,c)>s_2(b,c)
\end{eqnarray}
which, together with \eqref{PE-1}--\eqref{PE-2}, imply
\begin{equation}
\label{PEcase4} s_1(b,c)>s_3(b,c)>s_2(b,c)
\end{equation}
Suppose first that $s_2(b,c)>0$. Then, by \eqref{PEcase4} and 
\eqref{PE1}--\eqref{PE3}, 
$s_1(b,c)=3$, $s_2(b,c)=1$ and $s_3(b,c)=2$. From $s_1(b,c)=3$ and \eqref{PIF3} 
we also get 
$s_1(b,a)>0$ and $s_1(a,c)>0$. Hence, by \eqref{PE-1}, $s_2(b,a)>s_1(b,a)>0$ and, 
by 
\eqref{PEcase4c}, $s_2(a,c)>s_1(a,c)>0$. 
These inequalities and \eqref{PIF3} together imply $s_2(b,c)=3$, 
which is a contradiction.\\
Now suppose instead that $s_2(b,c)<0$, i.e. $s_2(c,b)>0$. 
It follows from \eqref{PEcase4} that $s_2(c,b)=3$, $s_3(c,b)=2$ and $s_1(c,b)=1$. 
Suppose $s_1(a,c)>0$. 
From \eqref{PEcase4c} and \eqref{PE1}--\eqref{PE3}, $s_2(a,c)>0$. 
Since $s_2(a,c)>0$ and $s_2(c,b)>0$, by \eqref{PIF3} we get 
$s_2(a,b)>s_2(c,b)=3$, 
which is impossible. Hence, $s_1(c,a)>0$ holds instead and, from 
\eqref{PEcase4c} and 
\eqref{PE1}--\eqref{PE3}, $s_1(c,a)>s_2(c,a)>0$ is also true. 
Suppose $s_2(a,b)>0$ holds too. By \eqref{PE-1}, $s_1(a,b)>0$. 
By \eqref{PIF3} and $s_1(c,a)>0$, $s_1(a,b)>0$ we get $s_1(c,b)=3$, 
a contradiction. 
Hence, $s_2(b,a)>0$ must be true instead and, by \eqref{PE-1}, 
$s_2(b,a)>s_1(b,a)>0$ 
also. 
So, we have $s_2(c,b)>0$ and $s_2(b,a)>0$, which, by \eqref{PIF3}, 
implies $s_2(c,a)>s_2(c,b)=3$. This too is a contradiction.

\noindent Hence, $w^4Dw^1$ is impossible for such $w^3$ and $w^4$ also.

Next, suppose $k=5$. Arguing as above, allocations $w^1,\ldots, w^4$ 
in \eqref{sequence} 
must be as in one of the four cases considered previously. 
Combined with the fact that each $w^i$ in sequence $(w^1,\ldots,w^5)$ 
must be distinct 
and the notational convention $w^1=(a,b,c)$ and $w^2=(b,a,c)$, 
this gives rise to the following four possibilities:
$$
\begin{array}{lll}
w^3=(b,c,a), & w^4=(c,b,a), & w^5=(c,a,b)\\
w^3=(c,a,b), & w^4=(c,b,a), & w^5=(b,c,a)\\
w^3=(b,c,a), & w^4=(a,c,b), & w^5=(c,a,b)\\
w^3=(c,a,b), & w^4=(a,c,b), & w^5=(b,c,a)\\
\end{array}
$$

Clearly, because either $w^5=(b,c,a)$ or $w^5=(c,a,b)$ must hold
in all four cases, 
and recalling that $w^1=(a,b,c)$, it cannot be that $w^5Dw^1$.

Finally, suppose $k=6$. With allocations $w^1,\ldots, w^5$ 
in \eqref{sequence} being as in the $k=5$ case 
that was just considered above, $w^6$ can only coincide with allocation $(a,c,b)$ 
in each of the four relevant cases. 
In view of the previous steps, these are as follows:\\
\noindent \textit{Case 1:} $w^1=(a,b,c)$, $w^2=(b,a,c)$, $w^3=(b,c,a)$, 
$w^4=(c,b,a)$, $w^5=(c,a,b)$, $w^6=(a,c,b)$. 
By definition of $D$, and by the above assumptions: 
$w^1Dw^2 \Longrightarrow s_1(a,b)>s_2(a,b)$; 
$w^2Dw^3 \Longrightarrow s_2(a,c)>s_3(a,c)$;
$w^3Dw^4 \Longrightarrow s_1(b,c)>s_2(b,c)$;
$w^4Dw^5 \Longrightarrow s_3(a,b)>s_2(a,b)$;
$w^5Dw^6 \Longrightarrow s_2(a,c)>s_1(a,c)$; and 
$w^6Dw^1 \Longrightarrow s_3(b,c)>s_2(b,c)$.
It follows that

\vspace{-30pt}

\begin{equation}
\label{k=6.part1} s_2(a,c)>s_1(a,c)>s_1(a,b)>s_2(a,b)
\end{equation}

\vspace{-15pt}

Suppose $s_2(a,b)>0$. Then, \eqref{k=6.part1} implies $s_2(a,c)>3$, 
which contradicts 
canonicality. 
If $s_2(b,a)>0$ instead, then \eqref{k=6.part1} together with skew-symmetry of 
$s_2$ implies 
$s_2(b,a)>3$ and results in the same contradiction.\\
\noindent \textit{Case 2:} $w^1=(a,b,c)$, $w^2=(b,a,c)$, $w^3=(c,a,b)$, 
$w^4=(c,b,a)$, $w^5=(b,c,a)$, $w^6=(a,c,b)$. 
Notice that the postulated dominance implications
$w^1Dw^2\Longrightarrow s_1(a,b)>s_2(a,b)$, 
$w^3Dw^4\Longrightarrow s_2(a,b)>s_3(a,b)$ and 
$w^5Dw^6\Longrightarrow s_3(a,b)>s_1(a,b)$ 
lead to $s_1(a,b)>s_2(a,b)>s_3(a,b)>s_1(a,b)$, which is absurd.\\
\noindent \textit{Case 3:} $w^1=(a,b,c)$, $w^2=(b,a,c)$, $w^3=(b,c,a)$, 
$w^4=(a,c,b)$, 
$w^5=(c,a,b)$, $w^6=(a,c,b)$. 
Observe here that the postulated dominance implications
$w^4Dw^5 \Longrightarrow s_1(a,c)>s_2(a,c)$ and 
$w^5Dw^6 \Longrightarrow s_2(a,c)>s_1(a,c)$ 
directly contradict each other.\\
\noindent \textit{Case 4:} $w^1=(a,b,c)$, $w^2=(b,a,c)$, $w^3=(c,a,b)$, 
$w^4=(a,c,b)$, $w^5=(b,c,a)$, $w^6=(a,c,b)$. 
As in Case 2, the postulated dominance implications
$w^4Dw^5 \Longrightarrow s_1(a,b)>s_3(a,b)$ and 
$w^5Dw^6 \Longrightarrow s_3(a,b)>s_1(a,b)$ 
result in the same contradiction.

It has therefore been shown that $D$ is acyclic when $n=3$,
establishing the existence of an intensity-efficient allocation 
in this case. 
The potential non-existence when $n\geq 4$ is proved in the main text.
(cf Table \ref{tab:counterexample4}).\hfill $\blacksquare$\\

\noindent \textbf{\textit{Proof of Theorem \ref{thm:homeomonot-neces}.}}
Assume to the contrary that profile $S$, with the postulated 
partition structure, 
has a strictly monotonic but not homeo-monotonic 
set $\{N'\subseteq N,A'\subseteq A\}$.
Write $N':=\{1',2',\ldots,|N'|\}$
and $A':=\{a'_1,a'_2,\ldots,a'_{|A'|}\}$.
By definition, $|N'|=|A'|$ and 
$\succ_i^{A'}$ $=$ $\succ_m^{A'}$ $:=$ $\succ^{A'}$
for all $i,m\in N'$.
Without loss of generality, define this common 
$A'$-restricted preference by
$a'_{1} \succ^{A'} a'_{2} 
\succ^{A'} \cdots \succ^{A'} a'_{|A'|}$
(in the general case, subscripts will be indexed
by some permutation on $\{1,\ldots,|A'|\}$).
By strict monotonicity of $N'$ with respect to $A'$, 
the binary relation $\trianglerighteq'$ on $N'$ defined by 
$i\trianglerighteq'm$ $\Leftrightarrow$ 
$s'_i(a'_{j},a'_{j+1})>
s'_m(a'_{j},a'_{j+1})$ for all $j\leq |A'|-1$
is a linear order.

From Lemma \ref{lem:pareto-necessary}, whose assumption is 
satisfied here, we know that every Pareto-efficient allocation 
$x$ must allocate all items in $A'$ to agents in $N'$. 
Furthermore, the way in which this is done is irrelevant:
if $x$ is a Pareto-efficient allocation, then so is 
allocation $x'$ that is 
otherwise identical to $x$ except that it reassigns items 
in $A'$ to agents in $N'$ under some permutation.

We will show that for every Pareto-efficient allocation
there is another such allocation that intensity-dominates it;
hence, that no intensity-efficient allocation exists.
To this end, let $x$ be any Pareto-efficient 
allocation that assigns the items in $A'$ to the agents
in $N'$ according to the $\trianglerighteq'$ ordering: 
$a'_1$ to $1'$; $a'_2$ to $2'$; $\ldots$; $a'_{|A'|}$ to $|N'|$.
By the strict monotonicity of $\{N',A'\}$, 
which is reflected in $\trianglerighteq'$, we have
$$
\begin{array}{lllllllll}
s'_1(a'_{i},a'_{i+1}) & > & s'_2(a'_{i},a'_{i+1}) & > &
\cdots & > & s'_{|N'|}(a'_{i},a'_{i+1}) && \text{for all } i\leq |A'|-1
\end{array}
$$
Since $\{N',A'\}$ is strictly monotonic but 
not homeo-monotonic, there are $j,l\in N'$ 
such that $j\trianglerighteq'l$, $l-j>1$ and 
$s'_l(a'_j,a'_l)>s'_j(a'_j,a'_l)$; that is 
the $\trianglerighteq'$ order is violated for agents $j,l$ and
non-consecutively-ranked items $a'_j,a'_l$.
Since, by assumption, $(x_j,x_l)=(a'_j,a'_l)$, this implies that 
the allocation $x'$ which is identical to $x'$ except
that $(x'_j,x'_l)=(a'_l,a'_j)$ intensity-dominates $x$.

Now define $x''$ from $x'$ by making the single item swap between 
agents $l-1$ and $j$ (note that, from the above assumptions, 
$j\trianglerighteq'l-1\trianglerighteq'l$ is true). 
Since $j\trianglerighteq' l-1$, and, by construction, 
$x_{l-1}=x'_{l-1}=a'_{l-1}$ and $x'_j=a'_l$ are also true, 
this leads to $s'_j(a'_{l-1},a'_l)\equiv 
s'_j(x''_{l-1},x''_j)>s'_{l-1}(x'_{l-1},x'_j)\equiv 
s'_{l-1}(a'_{l-1},a'_l)$. 
Thus, $x''$ intensity-dominates $x'$, which contradicts 
the above postulate. 
Now construct $x'''$ by swapping items $a'_l$ and $a'_{l-1}$,
keeping all other assignments in $x''$ the same.
That is, $x'''$ and $x''$ differ only in that the former 
assigns $a'_l$ to agent $l$ and $a'_{l-1}$ to $j$. 
By strict monotonicity and $j\trianglerighteq'l$  
we have $s_j(a'_{l-1},a'_l)\equiv s_j(x'''_j,x'''_l)>
s_l(x''_l,x''_j)\equiv s_l(a'_{l-1},a'_l)$.
Thus, $x'''$ intensity-dominates $x''$.
Now notice that $x$ and $x'''$ are identical except that
they swap items $a'_j$ and $a'_{l-1}$ between agents 
$j$ and $l-1$. That is, $(x'''_{l-1},x'''_j)=(a'_j,a'_{l-1})=
(x_j,x_{l-1})$. 
By strict monotonicity and $j\trianglerighteq' l-1$
we have $s_j(x_j,x_{l-1})>s_{l-1}(x'''_{l-1},x'''_j)$.
That is, $x$ intensity-dominates $x'''$. 
We have therefore established the intensity-dominance cycle 
$x'''Dx''Dx'DxDx'''$ between these four Pareto-efficient 
allocations. 

Suppose, finally, that there exists a Pareto-efficient 
allocation $y$ that is distinct from the above four. 
In particular, 
$y\neq (a'_1,a'_2,\ldots,a'_{|A'|})\equiv x$. 
If $y$ features a pairwise swap of items relative to $x$, 
then, by strict monotonicity, $x$ intensity-dominates $y$.
Hence, $y$ and $x$ are incomparable by intensity dominance
(for example, $y$ is a clockwise rotation of $x$).
Now, since the assignment in $y$ violates the order $\trianglerighteq'$
that is defined by the postulated strict monotonicity of 
$\{N',A'\}$, the allocation $y'$ that is derived from $y$ by swapping items
between agents according to $\trianglerighteq'$ intensity-dominates
$y$. Applying this reasoning repeatedly, and since $|A'|=|N'|\leq n$,  
there is a finite sequence of allocations $y^1,y^2,\ldots,y^m$
such that $y^1=y$, $y^m=x$ and $y^{i+1}$ intensity-dominates 
$y^i$ for all $i\leq m-1$. 

We have therefore shown that every Pareto-efficient allocation 
at $S$ is intensity-dominated by another such allocation. 
Hence, $S$ does not have an intensity-efficient allocation.
\hfill $\blacksquare$\\

\noindent \textbf{\textit{Proof of Proposition \ref{prp:complexity}.}}
Let $x$ be the Pareto-efficient allocation that is found in Step 2 
under the RP ordering $\gg_1$ that was drawn in the algorithm's 
first step. 
In the worst-case scenario, Step 4 always---with 
the exception of the agent 
ranked second-last by the ordering $\gg_2$ drawn in Step 3---performs 
exhaustive search and removes one agent and one object in every 
iteration: 
Starting with $1{_{\gg_2}},x_1{_{\gg_2}}$, after $n-1$ 
searches it removes this pair.
Continuing with $2{_{\gg_2}},x_2{_{\gg_2}}$, 
after $n-2$ searches it removes that pair.
$\cdots$ 
Continuing with $(n-1){_{\gg_2}},x_{(n-1)}{_{\gg_2}}$, 
after one search it removes both 
$(n-1){_{\gg_2}},x_{(n-1)}{_{\gg_2}}$ and 
$n{_{\gg_2}},x_n{_{\gg_2}}$, either by implementing  the 
swap $x_{(n-1)}{_{\gg_2}}\leftrightarrow x_n{_{\gg_2}}$ or otherwise.
The total number of searches in the worst-case scenario is therefore 
$$
\begin{array}{llclcllll}
\sum\limits_{m=1}^{n-1}(n-m) & = 
& n^2-n-\sum\limits_{m=1}^{n-1}m & = & n^2-\sum\limits_{m=1}^{n}m & = 
& & & \\
& & & & & & \\
& & n^2-\dfrac{n}{2}(n+1)& = & \dfrac{1}{2}n(n-1) & = & \displaystyle{{n\choose 2}} 
& &  
\end{array}
$$
Given that a Pareto-efficient allocation can be found by 
RP in Step 2 in $O(n)$ time \cite[Section 6.2]{manlove13}, 
it follows that the algorithm terminates in at most 
${{n}\choose{2}}$ steps, hence in $O(n^2)$ time.

It remains to be shown that, if the final allocation 
is $y\neq x$, then $yDx$.
To this end, let $y\neq x$ be the final allocation.
By construction, $x$ and $y$ 
differ \textit{only} by pairwise swaps,  
with Step 4 ensuring that a swap 
$$
\underbrace{(b,a)}_{x} \mapsto \underbrace{(a,b)}_{y}
$$
is implemented between two agents only if some 
agent $i$ who is ranked higher in 
$\gg_2$ than another agent $j$ prefers $a$ to $b$ more than 
the latter agent does, i.e. only if 
$s_i(a,b)\equiv s_i(x_j,x_i)>s_j(x_j,x_i)$ for such 
agents $i$ and $j$.
Since, by construction, no swap takes place unless it is 
in the intensity-improving direction, and since $x$ and $y$
are otherwise the same because of the forced exit of all 
agents and items involved in such a swap, and of every agent 
$i$ and item $x_i$ for whom such a pair could not be constructed, 
it follows that $y\neq x$ implies $yDx$.
\hfill $\blacksquare$

\bibliography{allocations,welfare}
\bibliographystyle{ecta-fullname}

\end{document}